\documentclass[aip,jcp,reprint,amsmath,amssymb,floatfix,citeautoscript]{revtex4-1}
\usepackage{cancel}
\usepackage{amsmath}
\usepackage{mathtools}
\usepackage{physics}
\usepackage{graphicx}
\usepackage{amssymb}
\usepackage{amsthm}
\usepackage{bm}
\usepackage{dcolumn}
\usepackage{braket}
\usepackage{longtable}
\usepackage{ragged2e}
\usepackage{txfonts}
\usepackage[version=3]{mhchem} 
\usepackage[T1]{fontenc}       
\usepackage{pstricks}
\usepackage{siunitx}
\usepackage[colorlinks=true,allcolors=blue]{hyperref}
\usepackage[capitalise]{cleveref}
\usepackage{multirow}
\usepackage{algorithm}         
\usepackage[noend]{algpseudocode}     

\let\tr\undefined
\newcommand*\tr[1]{\mathrm{#1}}
\newcommand*\mat[1]{\mathbf{#1}}
\newcommand*\mc[1]{\mathcal{#1}}

\definecolor{blackpink}{RGB}{200, 64, 200}

\newcommand*\me{\mathrm{e}}
\newcommand*\md{\mathrm{d}}
\newcommand*\mi{\mathrm{i}}
\newcommand*\uvec[1]{\hat{\bm{#1}}}
\newcommand*\tphi{\tilde{\phi}}

\allowdisplaybreaks

\begin{document}

\title
{Tight distance-dependent estimators for screening two-center and three-center short-range Coulomb integrals over Gaussian basis functions}

\author{Hong-Zhou Ye}
\email{hzyechem@gmail.com}
\affiliation
{Department of Chemistry, Columbia University, New York, New York 10027, USA}
\author{Timothy C. Berkelbach}
\email{tim.berkelbach@gmail.com}
\affiliation
{Department of Chemistry, Columbia University, New York, New York 10027, USA}
\affiliation
{Center for Computational Quantum Physics, Flatiron Institute, New York, New York 10010, USA}

\begin{abstract}
    We derive distance-dependent estimators for two-center and three-center electron repulsion integrals over a short-range Coulomb potential, $\tr{erfc}(\omega r_{12})/r_{12}$.
    These estimators are much tighter than one based on the Schwarz inequality and can be viewed as a complement to the distance-dependent estimators for four-center short-range Coulomb integrals and for two-center and three-center full Coulomb integrals previously reported.
    Because the short-range Coulomb potential is commonly used in solid-state calculations, including those with the HSE functional and with our recently introduced range-separated periodic Gaussian density fitting, we test our estimators on a diverse set of periodic systems using a wide range of the range-separation parameter $\omega$.
    These tests demonstrate the robust tightness of our estimators, which are then used with integral screening to calculate periodic three-center short-range Coulomb integrals with linear scaling in system size.
\end{abstract}

\maketitle

\section{Introduction}
\label{sec:introduction}

    Accurate estimators of electron repulsion integrals (ERIs) over pairs of charge densities are essential ingredients for large-scale electronic structure calculations using Gaussian-type orbitals (GTOs).
    Screening based on these estimators can be used to avoid computing negligible integrals and thereby achieve reduced computational scaling in both Hartree-Fock (HF) Coulomb \cite{White94CPL,Strout95JCP,White96CPL,White96JCP,Strain96Science,Challacombe96JCP} and
    exchange \cite{Schwegler96JCP,Challacombe97JCP,Ochsenfeld98JCP} problems and
    electron correlation methods \cite{Schutz99MP,Lambrecht05JCP,Doser09JCP,Maurer13JCP}.
    Despite its simplicity and behavior as a rigorous upper bound, the well-known Schwarz inequality \cite{Dyczmons73TCA,Haser89JCC,Gill94CPL} does not capture the decay of ERIs with the distance between the charge densities. \cite{Lambrecht05JCP}
    This has led to the development of tight distance-dependent integral estimators for conventional four-center ERIs, \cite{Lambrecht05JCP,Doser09JCP,Maurer12JCP,Maurer13JCP} as well as two-center and three-center ERIs, \cite{Hollman15JCP,Valeev20JCP} which appear in many semi-empirical methods \cite{Peels20JCTC} and the density fitting method \cite{Whitten73JCP,Dunlap79JCP,Mintmire82PRA}.

    In addition to the bare Coulomb operator, other potentials commonly appear in the ERIs. \cite{Savin95IJQC,Leininger97CPL,Adamson99JCC,Iikura01JCP,Heyd03JCP,Toulouse04PRA,Yanai04CPL,Refaely13PRB,Lutsker15JCP}
    One of the most widely used is the Coulomb potential attenuated by the complementary error function,
    \begin{equation}    \label{eq:sr_coulomb_pot}
        g_{\omega}(r_{12})
            = \frac{\tr{erfc}(\omega r_{12})}{r_{12}},
    \end{equation}
    which we henceforth refer to as the \emph{short-range} (SR) Coulomb potential.
    The SR Coulomb potential reduces to the full Coulomb potential for $\omega \to 0$ and $\delta(r_{12})$ for $\omega \to \infty$, thus connecting the full ERIs to the overlap integrals between two charge distributions. \cite{Jung05PNAS,Reine08JCP}
    The four-center SR ERIs are used in calculating the screened exchange energy in the Heyd-Scuseria-Ernzerhof (HSE) exchange correlation functional \cite{Heyd03JCP,Heyd06JCP}, whose application in solids is motivated by the unphysical behavior of long-range exchange in metals.
    Distance-dependent estimators for the four-center SR ERIs were first derived by Izmaylov and co-workers \cite{Izmaylov06JCP}, which have since been used for the efficient evaluation of the HSE exchange integrals \cite{Guidon08JCP,Guidon09JCTC,Shang11JCP,Beuerle17JCP}.

    The two-center and three-center SR ERIs were first used in local density fitting for finite systems \cite{Jung05PNAS,Reine08JCP} and screening was done according to the Schwarz inequality \cite{Reine08JCP}, which is suboptimal, as discussed above.
    More recently, the two of us introduced a global density fitting scheme for periodic systems \cite{Ye21JCP} where the use of range separation, in the spirit of Ewald summation, results in the appearance of two-center and three-center SR ERIs, which has motivated us to find tight estimators for integral screening.
    To the best of our knowledge, there has been no systematic studies on the estimators for two-center and three-center SR ERIs, which we aim to address in this work.

    Although our estimators are expected to work equally well in both finite and periodic calculations, we choose periodic systems in this work to demonstrate the practical use of the estimators.
    We develop algorithms for efficiently evaluating the periodic two-center and three-center SR ERIs, where the estimators are used to truncate the infinite lattice sum and avoid the calculation of unimportant integrals.
    We show that highly controlled accuracy of the computed periodic integrals can be achieved over a wide range of $\omega$ values.
    We analyze how the computational scaling of the lattice sum changes with $\omega$ and show that the computational cost scales linearly with the system size.

    This paper is organized as follows.
    In \cref{subsec:notations}, we establish our notation, and in \cref{subsec:j2cest,subsec:j3cest,subsec:pGTO_to_cGTO}, we present the derivation of the distance-dependent estimators for two-center and three-center SR ERIs, first for primitive GTOs and then extended to contracted GTOs.
    In \cref{subsec:pbcj23c}, we describe our algorithms for efficiently computing the periodic two-center and three-center SR ERIs, where the estimators derived in previous sections play the key role to truncate the infinite lattice sum and perform integral screening.
    After giving computational details in \cref{sec:computational_details}, we present numerical data in \cref{sec:results_and_discussions} to assess the tightness and accuracy of our estimators.
    We also discuss the favorable computational scaling for the lattice sum enabled by using the estimators.
    In \cref{sec:concluding_remarks}, we conclude by pointing out a few future directions.

\section{Theory}
\label{sec:theory}


    \subsection{Notations}
    \label{subsec:notations}

    In this work, a primitive GTO (pGTO) with principal angular momentum $l$, projected angular momentum $m$, and Gaussian exponent $\zeta_a$ is defined as \cite{Schlegel95IJQC}
    \begin{equation}    \label{eq:pGTO_def}
        \phi_{alm}(\bm{r})
            = N_{al} r^{l} y_{lm}(\uvec{r}) \me^{-\zeta_a r^2}
    \end{equation}
    where
    \begin{equation}    \label{eq:pGTO_normcoeff}
        N_{al}
            = \sqrt{\frac{2(2\zeta_a)^{l+3/2}}{\Gamma(l+3/2)}}
    \end{equation}
    is the radial normalization factor, and
    \begin{equation}    \label{eq:solid_ylm}
    \begin{split}
        y_{lm}(\theta,\varphi)
            &= \sqrt{\frac{2l+1}{4\pi} \frac{(l+|m|)!}{(l-|m|)!}
            (2 - \delta_{m0})} \\
            &\qquad \times P_{l}^{|m|}(\cos\theta)
            \left\{
            \begin{split}
                \cos(m\varphi)\quad& m \geq 0,  \\
                \sin(|m|\varphi)\quad& m < 0.
            \end{split}
            \right.
    \end{split}
    \end{equation}
    is the angular part of a real solid harmonic function.
    A contracted GTO (cGTO) is a linear combination of a group of concentric pGTOs that have the same angular momentum but differ in their Gaussian exponents,
    \begin{equation}    \label{eq:cGTO_def}
        \chi_{\mu}(\bm{r}; \bm{R})
            = \sum_{a} C_{a \mu} \phi_{al_{\mu}m_{\mu}}(\bm{r}-\bm{R}).
    \end{equation}
    A shell refers to a set of GTOs differing only by the projected angular momentum $m$.
    There are $2l+1$ orbitals in a shell of angular momentum $l$.
    Throughout this work, we consider atomic orbital (AO) basis sets that contain both primitive and contracted GTOs and auxiliary basis sets that are all primitive GTOs.
    Unless otherwise stated, the two-center ERIs are over two auxiliary orbitals, and the three-center ERIs are over the product of two AOs (in the bra) and an auxiliary orbital (in the ket).

    In the derivation below, we omit the labels for angular momentum and use $\phi_a \equiv \phi_{a l_a m_a}$ in cases without possible confusion.
    We also omit the radial normalization factor $N_{al}$, which is multiplicative and can be readily recovered if necessary.

    \subsection{Two-center SR ERIs over pGTOs}
    \label{subsec:j2cest}

    Consider the SR ERIs over two pGTOs
    \begin{equation}    \label{eq:j2c_def}
        J_{ab}(\bm{R})
            = \int \md\bm{r}_1 \int \md\bm{r}_2\,
            \phi_{a}(\bm{r}_1-\bm{R}) \phi_b(\bm{r}_2) g_{\omega}(r_{12})
    \end{equation}
    where we choose a coordinate system where the ket orbital is centered at the origin. The ERI $J_{ab}(\bm{R})$ may take $(2l_a+1) \times (2l_b+1)$ possible values $\{J_{ab}^{m_am_b}(\bm{R})\}$, each corresponding to a specific choice of $(m_a, m_b)$.
    Our goal in this section is to derive an approximate formula for estimating the shell-wise Frobenius norm
    \begin{equation}    \label{eq:j2c_Fnorm}
        \|J_{ab}^{m_am_b}(\bm{R})\|_{\tr{F}} \equiv
        \mc{J}_{ab}(R)
            = \bigg\{
                \sum_{m_a=-l_a}^{l_a}\sum_{m_b=-l_b}^{l_b}
                \big[
                    J_{ab}^{m_am_b}(\bm{R})
                \big]^2
            \bigg\}^{1/2}
    \end{equation}
    which only depends on the distance $R$ between the two orbitals due to the rotational invariance of the Frobenius norm.

    \subsubsection{The $O_0 v_0$ estimator}
    \label{subsubsec:O0v0}

    We begin by considering the simplest case of two $s$-type orbitals.
    The exact expression for \cref{eq:j2c_def} in this special case is well-known \cite{Izmaylov06JCP}
    \begin{equation}    \label{eq:j2c_ss}
        J_{a(l_a=0)b(l_b=0)}
            = \frac{O_{a0} O_{b0}}{R}
            \big[
                \tr{erfc}(\eta_{ab\omega}^{1/2} R) -
                \tr{erfc}(\eta_{ab}^{1/2} R)
            \big]
    \end{equation}
    where
    \begin{subequations}    \label{eq:etaab_etaabw_def}
    \begin{align}
        \eta_{ab}
            &= (\zeta_a^{-1}+\zeta_b^{-1})^{-1}, \\
        \eta_{ab\omega}
            &= (\eta_{ab}^{-1} + \omega^{-2})^{-1},
    \end{align}
    \end{subequations}
    and $O_{a0} = \pi/(2\zeta_a^{2/3})$ is the charge of $\phi_{a(l_a = 0)}$.
    Since the complementary error function decays exponentially with its argument and $\eta_{ab} > \eta_{ab\omega}$, only the first term of \cref{eq:j2c_ss} survives at large $R$.
    This leads to what we call the $O_0 v_0$ estimator
    \begin{equation}    \label{eq:j2cest_O0v0}
        \mc{J}^{O_0 v_0}_{ab}(R)
            = O_{a0} O_{c0} \frac{\tr{erfc}(\eta_{ab\omega}^{1/2}R)}{R},
    \end{equation}
    which can be applied for orbitals of arbitrary angular momenta.
    The name comes from interpreting \cref{eq:j2cest_O0v0} as two charges or zeroth-order multipoles (hence $O_0$) interacting via an \emph{effective} SR Coulomb potential
    \begin{equation}   \label{eq:v0_def}
        v_0(\eta_{ab\omega}, R)
            = g_{\eta_{ab\omega}^{1/2}}(R)
            = \frac{\tr{erfc}(\eta_{ab\omega}^{1/2}R)}{R}.
    \end{equation}
    Despite the formal similarity between the $O_0 v_0$ estimator and the classical Coulomb interaction between two point charges, we emphasize that the interpretation above is phenomenological rather than physical.
    As pointed out by Izmaylov and co-workers \cite{Izmaylov06JCP}, the fact that $v_0$ depends on the orbital exponents [i.e.,\ not simply $\tr{erfc}(\omega R)/R$] means that \cref{eq:j2cest_O0v0} is not a classical multipole interaction.
    Nonetheless, we will see below that the phenomenological interpretation applies for orbitals of higher angular momentum, too.

    The $O_0 v_0$ estimator shows no dependence on the orbital angular momentum $l$. One thus expects it to be accurate only for integrals over e.g.,\ $s$- and $p$-type orbitals.

    \subsubsection{The $O_l v_l$ estimator}
    \label{subsubsec:Olvl}

    Let us now consider the general case of \cref{eq:j2c_def} with arbitrary angular momenta, $l_a$ and $l_b$.
    The real-space double integral in \cref{eq:j2c_def} can be turned into a single integral in reciprocal space by using the Fourier transforms from \cref{app:fourier}; the result is
    \begin{equation}    \label{eq:j2c_qint}
    \begin{split}
        J_{ab}^{m_am_b}(\bm{R})
            &= \mi^{l_{ab}} 2^{-l_{ab}+1}
            \pi^{-1} \gamma_{al_a} \gamma_{bl_b} \times \\
            &\quad{} \sum_{l=|l_a-l_b|}^{l_{ab}} \sum_{m=-l}^{l} \mi^{l} T_{l_am_a,l_bm_b}^{lm} y_{lm}(\hat{\bm{R}}) \times \\
            &\quad{} \big[
                I(R; l_{ab}, l, \eta_{ab}) - I(R; l_{ab}, l, \eta_{ab\omega})
            \big]
    \end{split}
    \end{equation}
    where $l_{ab} = l_a + l_b$, $\gamma_{al} = (\pi/\zeta_a)^{3/2}\zeta_a^{-l}$,
    \begin{equation}    \label{eq:Tcoeff_def}
        T_{l_am_a,l_cm_c}^{lm}
            = \int \md\hat{\bm{q}}\,
            y_{lm}(\hat{\bm{q}})
            y_{l_am_a}(\hat{\bm{q}}) y_{l_cm_c}(\hat{\bm{q}})
    \end{equation}
    is the coefficient for angular momentum coupling, and
    \begin{equation}    \label{eq:I_def}
        I(R; L, l, \eta)
            = \int_0^{\infty}\md q\,
            q^{L} \me^{-q^2/(4\eta)} j_{l}(q R),
    \end{equation}
    with $j_l(x)$ the spherical Bessel function.

    To make progress, we show in \cref{app:asymptotic} that the contribution from \cref{eq:I_def} to \cref{eq:j2c_qint} is \emph{asymptotically} $l$-independent.
    As a result, we evaluate \cref{eq:I_def} with a convenient choice, $l = L$, and obtain a simple, closed-form expression
    \begin{equation}    \label{eq:I_LL_Gamma}
        I(R; L, L, \eta)
            = 2^{L-1} \pi^{1/2} R^{-(L+1)} \Gamma(L+1/2, \eta R^2),
    \end{equation}
    where $\Gamma(x,y)$ is the upper incomplete gamma function.
    With this result, in the large $R$ limit, \cref{eq:j2c_qint} simplifies to
    \begin{equation}    \label{eq:j2cest_mamb_gamma1}
        J_{ab}^{m_am_b}(\bm{R})
            \approx \frac{\gamma_{al_a}\gamma_{bl_b}}{\pi^{1/2}R^{l_{ab}+1}}
                \Gamma(l_{ab}+1/2,\eta_{ab\omega}R^2)
            \Theta_{l_a m_a, l_b m_b}^{lm}(\hat{\bm{R}})
    \end{equation}
    where
    \begin{equation}
        \Theta_{l_a m_a, l_b m_b}^{lm}(\hat{\bm{R}})
            = \mi^{l_{ab}} \sum_{l=|l_a-l_b|}^{l_{ab}} \sum_{m=-l}^{l} \mi^{l}
            T_{l_am_a,l_bm_b}^{lm} y_{lm}(\hat{\bm{R}})
    \end{equation}
    collects the angular dependence on $\bm{R}$.
    For estimating the Frobenius norm, $\mc{J}_{ab}(R)$, it is sufficient to use
    \begin{equation}
        \|\Theta_{l_am_a, l_bm_b}^{lm}(\hat{\bm{R}})\|_{\tr{F}}
            \approx \frac{\sqrt{(2l_a+1)(2l_b+1)}}{4\pi}.
    \end{equation}
    This simplifies \cref{eq:j2cest_mamb_gamma1} to what we call the $O_l v_l$ estimator
    \begin{equation}    \label{eq:j2cest_Olvl}
        \mc{J}_{ab}^{O_lv_l}(R)
            = O_{al_a} O_{bl_b}
            \frac{\Gamma(l_{ab}+1/2, \eta_{ab\omega} R^2)}
            {\pi^{1/2} R^{l_{ab}+1}},
    \end{equation}
    where
    \begin{equation}    \label{eq:Oal}
        O_{al}
            = \frac{\pi\sqrt{2l+1}}{2 \zeta_a^{l+3/2}}
    \end{equation}
    is the orbital multipole of arbitrary angular momentum.
    As for the $O_0 v_0$ estimator, the name $O_l v_l$ comes from interpreting \cref{eq:j2cest_Olvl} phenomenologically as two multipoles (hence $O_l$) interacting via an effective potential
    \begin{equation}    \label{eq:vl_def}
        v_{l_{ab}}(\eta_{ab\omega}, R)
            = \frac{\Gamma(l_{ab}+1/2, \eta_{ab\omega} R^2)}
            {\pi^{1/2} R^{l_{ab}+1}}.
    \end{equation}
    For $l_a = l_b = 0$, both the orbital multipole \cref{eq:Oal} and the effective potential \cref{eq:vl_def} reduce to their counterpart in \cref{subsubsec:O0v0}.
    Thus, the $O_l v_l$ estimator \cref{eq:j2cest_Olvl} is a direct generalization of the $O_0 v_0$ estimator to arbitrary orbital angular momenta.
    \Cref{eq:j2cest_Olvl} also parallels the estimator for two-center Coulomb ERIs obtained by Valeev and Shiozaki based on a multipole analysis \cite{Valeev20JCP}.

    In addition to $O_0 v_0$ and $O_l v_l$, one can readily write down two other estimators
    \begin{equation}    \label{eq:j2cest_Olv0}
        \mc{J}_{ab}^{O_lv_0}(R)
            = O_{al_a} O_{bl_b}
            v_0(\eta_{ab\omega}, R),
    \end{equation}
    \begin{equation}    \label{eq:j2cest_O0vl}
        \mc{J}_{ab}^{O_0v_l}(R)
            = O_{a0} O_{b0}
            v_{l_{ab}}(\eta_{ab\omega}, R)
    \end{equation}
    which consider the $l$-dependence for the orbital multipoles alone ($O_l v_0$) or the effective potential alone ($O_0 v_l$), respectively.
    In \cref{subsec:acc_j2cest}, we will see that only the $O_l v_l$ estimator is tight in all cases, and the comparison with the other three estimators help understand the importance of a correct treatment of the orbital angular momenta.

    \subsection{Three-center SR ERIs over pGTOs}
    \label{subsec:j3cest}

    Consider the SR ERIs over three pGTOs
    \begin{equation}    \label{eq:j3c_def}
    \begin{split}
        J_{abc}(\bm{R}_a,\bm{R}_b)
            &= \int\md\bm{r}_1\int\md\bm{r}_2\,
            \phi_{a}(\bm{r}_{1a}) \phi_{b}(\bm{r}_{1b})
            \phi_{c}(\bm{r}_2) g_{\omega}(r_{12})
    \end{split}
    \end{equation}
    where $\bm{r}_{1a} = \bm{r}_1 - \bm{R}_a$, $\bm{r}_{1b} = \bm{r}_1 - \bm{R}_b$, and we choose a coordinate system where the ket orbital is centered at the origin.
    The ERI $J_{abc}(\bm{R}_a,\bm{R}_b)$ can take $(2l_a+1) \times (2l_b+1) \times (2l_c+1)$ possible values $\{J_{abc}^{m_am_bm_c}(\bm{R}_a,\bm{R}_b)\}$, each corresponding to a specific choice of $(m_a, m_b, m_c)$.
    Our goal is again to derive an approximate formula for estimating the shell-wise Frobenius norm,
    \begin{equation}    \label{eq:j3c_Fnorm}
        \mc{J}_{abc}(\bm{R}, \bm{d})
            = \bigg\{
                \sum_{m_a=-l_a}^{l_a}\sum_{m_b=-l_b}^{l_b}\sum_{m_c=-l_c}^{l_c}
                \big[
                    J_{abc}^{m_am_bm_c}(\bm{R}_a,\bm{R}_b)
                \big]^2
            \bigg\}^{1/2},
    \end{equation}
    where on the left side of the equation we have switched to the representation of the bra separation, $\bm{d} = \bm{R}_b - \bm{R}_a$, and the bra-ket separation, $\bm{R} = (\zeta_a \bm{R}_a + \zeta_b \bm{R}_b) / \zeta_{ab}$, where $\zeta_{ab} = \zeta_a + \zeta_b$.
    Unlike the two-center case [\cref{eq:j2c_Fnorm}], $\mc{J}_{abc}(\bm{R},\bm{d})$ in general depends on both the norm and the orientation of the relevant position vectors.

    Our starting point for deriving an estimator for \cref{eq:j3c_Fnorm} is to turn the bra product distribution, $\rho_{ab}(\bm{r}_1) = \phi_a(\bm{r}_{1a})\phi_b(\bm{r}_{1b})$, into a sum of individual pGTOs,
    which will then reduce a three-center ERI into a sum of two-center ERIs, for which the $O_l v_l$ estimator~(\ref{eq:j2cest_Olvl}) is a good approximation.

    \subsubsection{The ISF estimator}
    \label{subsubsec:ISF}

    We begin with the simple case, $l_a = l_b = 0$, where the Boys relation~\cite{Boys50PRSLA,Szabo96Book} can be used,
    \begin{equation}    \label{eq:boys_relation}
        \phi_{a(l_a=0)}(\bm{r}_{1a})
        \phi_{b(l_b=0)}(\bm{r}_{1b})
            = y_{00} \me^{-\eta_{ab} d^2} \phi_{\zeta_{ab}(l=0)}(\bm{r}_{1\bm{R}}),
    \end{equation}
    which expresses the well-known result that the product of two $s$-type pGTOs is another $s$-type pGTO with an exponent $\zeta_{ab}$, located at the charge center $\bm{R}$, and scaled in magnitude by $y_{00} \me^{-\eta_{ab} d^2}$.
    Using this identity, a three-center ERI with two $s$-type bra orbitals and an arbitrary ket orbital is
    \begin{equation}
        J_{a(l_a=0)b(l_b=0)c}(\bm{R}, \bm{d})
            = \frac{\me^{-\eta_{ab}d^2}}{2\pi^{1/2}}
            J_{\zeta_{ab}(l_{ab}=0)c}(\bm{R}).
    \end{equation}
    Approximating $J_{\zeta_{ab}(l_{ab}=0)c}(\bm{R})$ by the $O_lv_l$ estimator~(\ref{eq:j2cest_Olvl}) leads to what we call the ISF estimator for three-center SR ERIs
    \begin{equation}    \label{eq:j3cest_ISF}
        \mc{J}_{abc}^{\tr{ISF}}(R, d)
            = \frac{\me^{-\eta_{ab}d^2}}{2\pi^{1/2}} O_{\zeta_{ab}0} O_{cl_c}
            v_{l_{c}}(\eta_{abc\omega}, R)
    \end{equation}
    where
    \begin{equation}    \label{eq:etaabcw_def}
        \eta_{abc\omega}
            = (\eta_{ab}^{-1}+\zeta_{c}^{-1}+\omega^{-2})^{-1}.
    \end{equation}
    The name comes from the fact that \cref{eq:j3cest_ISF} can be viewed as a direct extension of the work by Izmaylov, Scuseria, and Frisch \cite{Izmaylov06JCP} (ISF), who derived the exact expression for a four-center SR ERI over all $s$-type pGTOs using the Boys relation~(\ref{eq:boys_relation}) and then used it as an estimator for orbitals of arbitrary angular momentum.
    Here in \cref{eq:j3cest_ISF}, we include the $l$-dependence for the ket orbital via our two-center estimator~(\ref{eq:j2cest_Olvl}).

    Note that the ISF estimator (and all other three-center estimators derived below) depends on only the norm of $\bm{R}$ and $\bm{d}$.
    The lack of angular dependence here is not a crucial issue as applications such as integral screening are nearly exclusive to medium to large bra-ket separation $R$, where the effect of the orbital orientation is relatively weak.

    \subsubsection{The ISF$Q_0$ estimator}
    \label{subsubsec:ISFQ0}

    For non-$s$-type orbitals, the bra product is in general a sum of $l_{ab}+1$ terms according to the Gaussian product theorem (GPT) \cite{Besalu11JMC,Fermann20arXiv},
    \begin{equation}    \label{eq:GPT}
        \phi_{a}(\bm{r}_{1a})\phi_{b}(\bm{r}_{1b})
            = \me^{-\eta_{ab}d^2} \sum_{l=0}^{l_{ab}}
            \sum_{m=-l}^{l}
            L_{l_am_a,l_bm_b}^{lm}(\bm{d},\zeta_a,\zeta_b)
            \phi_{\zeta_{ab}(l)}(\bm{r}_{1\bm{R}}),
    \end{equation}
    where $\{L_{l_am_a,l_bm_b}^{lm}\}$ are related to the Talmi coefficients whose explicit expression can be derived in various ways. \cite{Matsuoka98JMST,Matsuoka98JCP}
    \Cref{eq:GPT} can also be viewed as expanding a product distribution by its multipole components, and the ISF estimator~(\ref{eq:j3cest_ISF}) keeps only the lowest-order multipole, i.e.,\ the charge, of the bra product distribution.

    One way to include the effect of higher-order multipoles is using the Schwarz $Q$-integral \cite{Haser89JCC} generalized for a SR Coulomb potential
    \begin{equation}
        \mc{Q}_{ab}(d)
            = \|V_{abab}(\bm{d})\|_{\tr{F}}^{1/2}
    \end{equation}
    where
    \begin{equation}
    \begin{split}
        V_{abab}(\bm{d})
            &= \int \md\bm{r}_1\int\md\bm{r}_2\,
            \phi_{a}(\bm{r}_1)\phi_{b}(\bm{r}_1-\bm{d}) \\
        &\hspace{6em} \times
            \phi_{a}(\bm{r}_2)\phi_{b}(\bm{r}_2-\bm{d})
            g_{\omega}(r_{12})
    \end{split}
    \end{equation}
    and the Frobenius norm is again shell-wise
    \begin{equation}
        \|V_{abab}(\bm{d})\|_{\tr{F}}
            = \bigg\{
                \sum_{m_{a}=-l_a}^{l_a}\sum_{m_{b}=-l_b}^{l_b}
                \sum_{m_{a}'=-l_a}^{l_a}\sum_{m_{b}'=-l_b}^{l_b}
                [V_{abab}^{m_am_bm_a'm_b'}(\bm{d})]^2
            \bigg\}^{1/2}.
    \end{equation}
    For the simplest case of two $s$-type pGTOs,
    \begin{equation}    \label{eq:Qab_ss}
        \mc{Q}_{a(l_a=0)b(l_b=0)}(d)
            = \me^{-\eta_{ab}d^2} \frac{\sqrt{2}}{\pi^{5/4}}
            \big(
                \eta_{abab}^{1/2} - \eta_{abab\omega}^{1/2}
            \big)^{1/2}
    \end{equation}
    where $\eta_{abab} = (\zeta_{ab}^{-1} + \zeta_{ab}^{-1})^{-1} = \zeta_{ab}/2$ and $\eta_{abab\omega} = (\eta_{abab}^{-1} + \omega^{-2})^{-1}$.
    By rewriting the exponential factor $\me^{-\eta_{ab}d^2}$ in \cref{eq:j3cest_ISF} using \cref{eq:Qab_ss}, we obtain what we call the ISF$Q_0$ estimator
    \begin{equation}    \label{eq:j3cest_ISFQ0}
        \mc{J}_{abc}^{\tr{ISF}Q_0}(R, d)
            = \frac{\pi^{3/4} \mc{Q}_{ab}(d) O_{cl_c}}
            {\big[ 2 (\eta_{abab}^{1/2} - \eta_{abab\omega}^{1/2}) \big]^{1/2}}
            v_{l_c}(\eta_{abc\omega}, R)
    \end{equation}
    We note that \cref{eq:j3cest_ISFQ0} parallels the $QVl$ estimator obtained by Hollman \textit{et al.}\ for three-center Coulomb ERIs. \cite{Hollman15JCP}

    \subsubsection{The ISF$Q_l$ estimator}
    \label{subsubsec:ISFQl}

    The ISF$Q_0$ estimator amounts to approximating the exact multipole expansion of the bra product distribution~(\ref{eq:GPT}) by the $l = 0$ term with \emph{a modified prefactor} to capture the overall effect of all higher-order terms.
    When $\omega$ is small, we expect this to be a good approximation, because the SR Coulomb potential resembles the full Coulomb potential, for which the classical multipole interaction, which decays faster for higher-order multipoles, is a good approximation \cite{Lambrecht05JCP,Maurer12JCP}.
    For large $\omega$, however, terms with $l > 0$ could be more important due to the incomplete gamma function in the effective potential~(\ref{eq:vl_def}).
    Using the ISF$Q_0$ estimator may cause underestimation in this regime.

    While it is possible to consider a full multipole expansion with approximate coefficients (\cref{subsubsec:ME}), a simpler amendment to the ISF$Q_0$ estimator is to restore the $l$-dependence and keep only the term of the maximum value.
    Specifically, we define the ISF$Q_l$ estimator as
    \begin{equation}    \label{eq:j3cest_ISFQl}
    \begin{split}
        \mc{J}^{\tr{ISF}Q_l}_{abc}(R,d)
            &= \frac{\pi^{3/4} \mc{Q}_{ab}(d) O_{cl_c}}{2^{1/2}} \\
            &\quad{}\times
            \max_{l} \bigg\{
                \big( \eta_{abab}^{l+1/2} -
                      \eta_{abab\omega}^{l+1/2} \big)^{-1/2}
                v_{l+l_c}(\eta_{abc\omega}, R)
             \bigg\},
    \end{split}
    \end{equation}
    where the maximization is over $\{0,1,\cdots,l_{ab}\}$ for $d > 0$ but $\{|l_a-l_b|,|l_a-l_b|+1,\cdots,l_{ab}\}$ for $d = 0$ by the properties of angular momentum coupling.
    We expect ISF$Q_l$ to essentially reduce to ISF$Q_0$ when $\omega$ is small, but corrects the underestimation of the latter for larger $\omega$.

    \subsubsection{The ME estimator}
    \label{subsubsec:ME}

    In principle, a more accurate account of the multipole expansion~(\ref{eq:GPT}) needs the GPT coefficients $\{L_{l_am_a,l_bm_b}^{lm}\}$.
    Consider a special case where both $\phi_{a}$ and $\phi_{b}$ are located on the $z$-axis and $m_{a} = m_{b} = 0$. In this case, the spherical GTO~(\ref{eq:pGTO_def}) becomes equivalent to a Cartesian GTO with $z$-component only,
    \begin{equation}
        \phi_{a(m_{a} = 0)}(\bm{r}_{1a})
            \sim z_{1a}^{l_a} \me^{-\zeta_a r_{1a}^2},
    \end{equation}
    and a similar expression holds for $\phi_{b(m_b=0)}(\bm{r}_{1b})$. The product distribution then becomes,
    \begin{equation}
        \phi_{a(m_a=0)}(\bm{r}_{1a}) \phi_{b(m_b=0)}(\bm{r}_{1b})
            \sim \me^{-\eta_{ab} d^2}
            \sum_{l=0}^{l_{ab}} L_{l_a,l_b}^{l}
            z_{1\bm{R}}^{l} \me^{-\zeta_{ab} r_{1\bm{R}}^2}
    \end{equation}
    where
    \begin{equation}    \label{eq:Talmi_coeff_zz}
        L_{l_a,l_b}^{l}
            = \sum_{l' = -l}^{l}{}^{'}
            {l_a \choose l_a'} {l_b \choose l_b'}
            d_{a}^{l_a - l_a'} d_{b}^{l_b - l_b'},
    \end{equation}
    gives the GPT coefficients in this special case, where $l_a' = (l+l')/2$, $l_b' = (l-l')/2$, $d_{a} = z_{\bm{R}} - z_a$, $d_{b} = z_{\bm{R}} - z_b$, and the primed summation means increment by $2$. Now for the general case where $\phi_a$ and $\phi_b$ are arbitrarily located and have arbitrary $(m_a, m_b)$, we can still use \cref{eq:Talmi_coeff_zz} to approximate the GPT coefficients if $d > 0$ and $d_{a}$ and $d_b$ are chosen to be
    \begin{equation}    \label{eq:dadb_1D}
    \begin{split}
        d_{a}
            &= -\|\bm{R}_{a}-\bm{R}\| = -(\zeta_b/\zeta_{ab}) d,    \\
        d_{b}
            &= \|\bm{R}_{b}-\bm{R}\| = (\zeta_a/\zeta_{ab}) d.
    \end{split}
    \end{equation}
    Using \cref{eq:Talmi_coeff_zz,eq:dadb_1D} leads to what we call the ME estimator (where ``ME'' stands for multipole expansion)
    \begin{equation}    \label{eq:j3cest_ME}
    \begin{split}
        \mc{J}_{abc}^{\tr{ME}}(R,d)
            &= \frac{\me^{-\eta_{ab} d^2} O_{c l_{c}}}{2\pi^{1/2}}
            \sum_{l=0}^{l_{ab}} |L_{l_a,l_b}^{l}(d,\zeta_a,\zeta_b)|
            O_{\zeta_{ab}l}
            v_{l+l_c}(\eta_{abc\omega}, R).
    \end{split}
    \end{equation}

    The case $d = 0$, i.e.,\ $\phi_a$ and $\phi_b$ are concentric, needs special consideration.
    As mentioned above, the GPT expansion in this case should range from $|l_a-l_b|$ to $l_{ab}$, while \cref{eq:Talmi_coeff_zz,eq:dadb_1D} predict all terms except for $l=l_{ab}$ vanish, which leads to a significant underestimation of the true integrals.
    To obtain a better approximation in this case, we assume a simple structure for the approximate GPT coefficients
    \begin{equation}    \label{eq:Talmi_coeff_d0}
        L_{l_a,l_b}^{l}(d=0, \zeta_a,\zeta_b)
            = \zeta_{ab}^{\alpha l + \beta l_{ab}} f(l_{ab},l)
    \end{equation}
    where $l = |l_a-l_b|, |l_a-l_b|+1, \cdots, l_{ab}$, and determine the parameters in \cref{eq:Talmi_coeff_d0} empirically from numerical tests.
    We found that the following choices work well
    \begin{equation}    \label{eq:Talmi_params_d0}
        \alpha = \frac{1}{2},
        \quad
        \beta = -\frac{1}{2},
        \quad
        f(l_{ab},l)
            = \bigg[\frac{(l_{ab}-1)!}{(l-1)!}\bigg]^{1/2}
    \end{equation}
    where we define $(-1)! = 1$.
    The numerical evidence for \cref{eq:Talmi_params_d0} is given in Figs.\ S1 and S2.
    \Cref{eq:Talmi_coeff_d0,eq:Talmi_params_d0} together with \cref{eq:Talmi_coeff_zz,eq:dadb_1D} thus complete the definition of the ME estimator~(\ref{eq:j3cest_ME}) for three-center SR ERIs.

    \subsection{From primitive to contracted GTOs}
    \label{subsec:pGTO_to_cGTO}

    The estimators derived above assume all orbitals are pGTOs [\cref{eq:pGTO_def}].
    To apply them for integrals over cGTOs [\cref{eq:cGTO_def}], we find that a one-term approximation works well in the appropriate large-$R$ limit. In this case, a cGTO, $\chi_{\mu}$, is replaced by its most diffuse pGTO component, $C_{\mu a^*}\phi_{a^*}$, and
    \begin{equation}    \label{eq:j3c_ctr_oneterm}
        \mc{J}_{\mu\nu c}(\bm{R},\bm{d})
            \approx C_{\mu a^*} C_{\nu b^*} \mc{J}_{ab c}(\bm{R}, \bm{d}),
    \end{equation}
    where $\mc{J}_{abc}(\bm{R},\bm{d})$ can be estimated by one of the three-center estimators derived in the previous section.
    For ISF$Q_0$~(\ref{eq:j3cest_ISFQ0}) and ISF$Q_l$~(\ref{eq:j3cest_ISFQl}), the Schwarz-$Q$ integrals can be calculated using the original cGTOs [i.e.,\ $\mc{Q}_{\mu\nu}(d)$] to effectively account for the contribution from other pGTOs in the cGTOs.


    \subsection{Periodic two-center and three-center SR ERIs with screening}
    \label{subsec:pbcj23c}

    As a practical application of the estimators derived above and also a means to test their accuracy, we show how to exploit these estimators to efficiently calculate periodic two-center and three-center SR ERIs.
    As mentioned in the introduction, these integrals are needed in our recently introduced periodic global density fitting scheme \cite{Ye21JCP} and would be needed in a density-fitted implementation of the HSE functional \cite{Heyd03JCP,Heyd06JCP}, among other possible applications.

    \subsubsection{Periodic two-center SR ERIs}
    \label{subsubsec:pbcj2c}

    A periodic system consists of a unit cell and its infinite periodic images, each specified by a lattice translational vector, $\bm{m}$, with $\bm{m}=\bm{0}$ the reference cell.
    Consider $n$ atom-centered GTOs, $\{\phi_a(\bm{r}-\bm{\tau}_a)\}$, in the reference cell, which, under the periodic boundary condition, become $n$ translationally adapted GTOs
    \begin{equation}    \label{eq:taGTO}
        \tphi_{a}^{\bm{k}}(\bm{r})
            = \sum_{\bm{m}} \me^{\mi \bm{k}\cdot\bm{m}}
            \phi_a(\bm{r}-\bm{m}_a)
    \end{equation}
    where $\bm{m}_a = \bm{m} + \bm{\tau}_a$ and $\bm{k}$ is a crystal momentum vector in the first Brillouin zone.
    In the following, we consider only the $\Gamma$-point Brillouin zone sampling with $\tphi_a \equiv \tphi_a^{\bm{k}=\bm{0}}$.
    This choice corresponds to an in-phase superposition of all cells and hence represents the most challenging case for integral screening.
    The method can be straightforwardly adapted to use other Brillouin zone sampling schemes.

    A periodic two-center SR ERI is
    \begin{equation}    \label{eq:pbcj2c_def}
        L_{ab}
            = \int_{\Omega}\md\bm{r}_1\int \md\bm{r}_2\,
            \tilde{\phi}_a(\bm{r}_1) \tilde{\phi}_b(\bm{r}_2)
            g_{\omega}(r_{12})
            = \sum_{\bm{m}} J_{ab}(\bm{m}_{ab_0})
    \end{equation}
    where $\Omega$ is the volume of a unit cell, $\bm{m}_{ab_0} = \bm{m}_{a}-\bm{\tau}_{b}$, and we used \cref{eq:taGTO,eq:j2c_def} to obtain the second equality, which is an infinite lattice sum.
    To calculate $L_{ab}$ to a finite precision $\epsilon$, we approximate the lattice sum by an integral over $\bm{R}_{ab} = \bm{m}_{ab_0}$ and analyze the error of truncating it at a finite $R_{ab}^{\tr{cut}}$
    \begin{equation}    \label{eq:trunc_err_Lab}
        \delta L_{ab}
            \sim \int_{R_{ab} > R_{ab}^{\tr{cut}}} \md\bm{R}_{ab}\,
            J_{ab}(\bm{R}_{ab})
            \sim f(\eta_{ab\omega}, R_{ab}^{\tr{cut}})
            \mc{J}_{ab}(R_{ab}^{\tr{cut}})
    \end{equation}
    where the form of the multiplicative prefactor $f(\eta_{ab\omega}, R_{ab}^{\tr{cut}})$ will be derived in \cref{subsubsec:latsum_prefac}.
    \Cref{eq:trunc_err_Lab} suggests the cutoff criterion for the lattice sum for $L_{ab}$,
    \begin{equation}    \label{eq:j2c_Rcut}
        f(\eta_{ab\omega}, R_{ab}^{\tr{cut}})
        \mc{J}_{ab}(R_{ab}^{\tr{cut}})
            = \epsilon,
    \end{equation}
    where $\mc{J}_{ab}(R)$ can be estimated using a two-center estimator from \cref{subsec:j2cest}.
    Once $R_{ab}^{\tr{cut}}$ is determined from \cref{eq:j2c_Rcut},
    \begin{equation}
        \mc{M}_{ab}
            = \{ \bm{m} \,|\, \|\bm{m}_{ab_0}\| \leq R_{ab}^{\tr{cut}}\}
    \end{equation}
    gives the set of cells needed for calculating $L_{ab}$ to the precision $\epsilon$ via the lattice summation~(\ref{eq:pbcj2c_def}).
    The union $\mc{M}_{a} = \cup_{b} \mc{M}_{ab}$ then gives all ``important'' cells for orbital $a$.
    An algorithm for efficiently calculating the entire $[\mat{L}]_{ab}$ matrix based on precomputed cutoffs $\{R_{ab}^{\tr{cut}}\}$ and cell data $\{\mc{M}_{a}\}$ is presented in \cref{alg:j2c}.

    \begin{algorithm}[H]
        \caption{\cref{eq:pbcj2c_def} by a truncated lattice sum.}
        \label{alg:j2c}
        \begin{algorithmic}
            \State Input: $\{R_{ab}^{\tr{cut}}\}$, $\{\mc{M}_{a}\}$
            \State Initialize: $[\mat{L}]_{ab} \gets $ zeros($n$,$n$)
            \For{$a$ in range($n$)}
                \For{$b$ in range($n$)}
                    \For{$\bm{m} \in \mc{M}_{a}$}
                        \If{$\|\bm{m}_{ab_0}\| \leq R^{\tr{cut}}_{ab}$}
                            \State $L_{ab} \mathrel{+}= J_{ab}(\bm{m}_{ab_0})$
                        \EndIf
                    \EndFor
                \EndFor
            \EndFor
        \end{algorithmic}
    \end{algorithm}

    \subsubsection{Periodic three-center SR ERIs}
    \label{subsubsec:pbcj3c}

    Let $\tphi$ and $\tilde{\varphi}$ denote two sets of translationally adapted GTOs [\cref{eq:taGTO}] of size $n_1$ and $n_2$, respectively (e.g., the sets of periodic AOs and auxiliary basis functions).
    A periodic three-center SR ERI (using $\Gamma$-point Brillouin zone sampling as discussed above) is
    \begin{equation}    \label{eq:pbcj3c_def}
    \begin{split}
        L_{abc}
            &= \int_{\Omega}\md\bm{r}_1 \int\md\bm{r}_2\,
            \tphi_a(\bm{r}_1)\tphi_b(\bm{r}_1)\tilde{\varphi}_c(\bm{r}_2)
            g_{\omega}(r_{12})  \\
            &= \sum_{\bm{m},\bm{n}} J_{abc}(\bm{m}_{ac_0},\bm{n}_{bc_0}),
    \end{split}
    \end{equation}
    where we used \cref{eq:j3c_def,eq:taGTO} to obtain the second equality, which is an infinite double lattice sum.
    The double lattice sum in \cref{eq:pbcj3c_def} can be rewritten to be over the bra separation, $\bm{d}_{ab} = \bm{m}_a - \bm{n}_b$, and the bra-ket separation, $\bm{R}_{abc} = \zeta_{ab}^{-1}(\zeta_{a}\bm{m}_{ac_0}+\zeta_{b}\bm{n}_{bc_0})$,
    \begin{equation}    \label{eq:pbcj3c_Rd}
        L_{abc}
            = \sum_{\bm{d}_{ab}} \sum_{\bm{R}_{abc}}
            J_{abc}(\bm{R}_{abc}, \bm{d}_{ab}).
    \end{equation}
    \Cref{eq:pbcj3c_Rd} is more convenient for truncation to compute $L_{abc}$ to a finite precision $\epsilon$ as we discuss now.

    First, the Schwarz inequality
    \begin{equation}    \label{eq:Schwarz_inequality}
        \mc{J}_{abc}(\bm{R}_{abc}, \bm{d}_{ab})
            \leq \mc{Q}_{ab}(d_{ab}) \mc{Q}_c
    \end{equation}
    with $\mc{Q}_c = [\mc{J}_{cc}(0)]^{1/2}$ gives the decay with $d_{ab}$ regardless of the value for $\bm{R}_{abc}$.
    We thus follow a similar derivation for \cref{eq:j2c_Rcut} and obtain an equation for the cutoff of $d_{ab}$,
    \begin{equation}    \label{eq:j3c_dcut}
        f(\eta_{ab}, d_{ab}^{\tr{cut}})
        \mc{Q}_{ab}(d_{ab}^{\tr{cut}}) \mc{Q}_c^{\tr{max}}
            = \epsilon
    \end{equation}
    where $\mc{Q}_c^{\tr{max}} = \max_c \mc{Q}_c$.

    Second, for a fixed $d_{ab} \leq d_{ab}^{\tr{cut}}$, \cref{eq:pbcj3c_Rd} reduces to a single lattice sum over $R_{abc}$.
    Following the same argument for obtaining \cref{eq:j2c_Rcut}, we determine a cutoff for $R_{abc}$ for a given bra separation $d_{ab}$ from solving
    \begin{equation}    \label{eq:j3c_Rcut}
        f(\eta_{abc\omega}, R_{abc}^{\tr{cut}}(d_{ab}))
        \mc{J}_{abc}(R_{abc}^{\tr{cut}}(d_{ab}), d_{ab})
            = \epsilon,
    \end{equation}
    where $\mc{J}_{abc}(R,d)$ can be estimated using one of the estimators from \cref{subsec:j3cest}.
    In principle, \cref{eq:j3c_Rcut} needs to be solved for all unique $d_{ab}$'s arising from all bra pairs $ab$ with $d_{ab} \leq d_{ab}^{\tr{cut}}$.
    This number could be very large, leading to high cost in both the CPU time and the storage.
    We avoid this difficulty by solving \cref{eq:j3c_Rcut} only for $d_{ab} \in \{d_{ab}^{(i)}\}_{i=1}^{n_{d_{ab}}}$,
    where $d_{ab}^{(i)} = (i-1)\Delta d$ for some chosen $\Delta d$ and $n_{d_{ab}} = \lfloor d_{ab}^{\tr{cut}}/\Delta d \rfloor$.
    The cutoff $R_{abc}^{\tr{cut}}(d_{ab}^{(i)})$ is then used for all bra pairs with $d_{ab} \in [d_{ab}^{(i)}, d_{ab}^{(i+1)})$.

    With these cutoffs,
    the double lattice sum for $L_{abc}$ contains only a finite number of terms given by $\mc{S}_{abc} = \cup_{i=1}^{n_{d_{ab}}} \mc{S}_{abc}^{(i)}$, where
    \begin{equation}    \label{eq:Sabc_i}
        \mc{S}_{abc}^{(i)}
            = \{
                (\bm{m},\bm{n}) \,|\,
                d_{ab}^{(i)} \leq d_{ab} < d_{ab}^{(i+1)},
                R_{abc} \leq R_{abc}^{\tr{cut}}(d_{ab}^{(i)})
            \}.
    \end{equation}
    The union $\mc{S}_{ab}^{(i)} = \cup_{c} \mc{S}_{abc}^{(i)}$ then gives the set of ``important'' cell pairs for a bra pair $ab$.
    An algorithm for efficiently calculating the entire $[\mat{L}]_{abc}$ tensor based on the cutoffs $\{R_{abc}^{\tr{cut}}\}$ and cell pair data $\{\mc{S}_{ab}^{(i)}\}$ is presented in \cref{alg:j3c}.

    \begin{algorithm}[H]
        \caption{\cref{eq:pbcj3c_def} by a truncated double lattice sum.}
        \label{alg:j3c}
        \begin{algorithmic}
            \State Input: $\{R_{abc}^{\tr{cut}}\}$, $\{\mc{S}_{ab}^{(i)}\}$
            \State Initialize: $[\mat{L}]_{abc} \gets $ zeros($n_1$,$n_1$,$n_2$)
            \For{$a$ in range($n_1$)}
                \For{$b$ in range($n_1$)}
                    \For{$i$ in range($n_{d_{ab}}$)}
                        \For{$(\bm{m},\bm{n})$ in $\mc{S}_{ab}^{(i)}$}
                            \For{$c$ in range($n_2$)}
                                \If{$R_{abc} \leq R_{abc}^{\tr{cut}}(d_{ab}^{(i)})$}
                                    \State $L_{abc} \mathrel{+}= J_{abc}(\bm{m}_{ac_0},\bm{n}_{bc_0})$
                                \EndIf
                            \EndFor
                        \EndFor
                    \EndFor
                \EndFor
            \EndFor
        \end{algorithmic}
    \end{algorithm}

    \subsubsection{An expression for the $f$-prefactor}
    \label{subsubsec:latsum_prefac}

    The three cutoffs discussed above all correspond to truncating a single lattice sum approximated by an integral to a finite precision $\epsilon$,
    \begin{equation}    \label{eq:trunc_err_gen}
        \delta
            \sim \int_{R > R_{\tr{cut}}} \md \bm{R}\, K(\bm{R})
            \lesssim \epsilon,
    \end{equation}
    where $K(\bm{R}) = J_{ab}(\bm{R})$ for \cref{eq:j2c_Rcut}, $\mc{Q}_{ab}(\bm{R})$ for \cref{eq:j3c_dcut},
    and $J_{abc}(\bm{R},\bm{d})$ with $\bm{d}$ fixed for \cref{eq:j3c_Rcut}.
    We approximate $J_{ab}$ and $J_{abc}$ by the corresponding distance-dependent estimators, which have the following asymptotic behavior at large $R$,
    \begin{equation}    \label{eq:KR_asymptotics}
        K(R)
            \sim R^{l-2} \me^{-\eta R^2},
    \end{equation}
    where $l = l_{ab}$ and $\eta = \eta_{ab\omega}$ for $J_{ab}$ and $l = l_{abc}$ and $\eta = \eta_{abc\omega}$ for $J_{abc}$, respectively.
    \Cref{eq:KR_asymptotics} also describes the asymptotics of $\mc{Q}_{ab}$ with $l = 0$ and $\eta = \eta_{ab}$ if we approximate it by $\mc{Q}_{a(l_a=0)b(l_b=0)}$ [\cref{eq:Qab_ss}].
    Combining \cref{eq:KR_asymptotics} with \cref{eq:trunc_err_gen} gives
    \begin{equation}    \label{eq:trunc_err_infRcutlimit}
        \delta
            \sim \frac{R_{\tr{cut}}^{l-1}}{\eta}
            \me^{-\eta R_{\tr{cut}}^2}
            \sim \frac{R_{\tr{cut}}}{\eta}
            K(R_{\tr{cut}}),
    \end{equation}
    which suggests that
    \begin{equation}    \label{eq:fprefac_theoretical}
        f(\eta, R_{\tr{cut}})
            = \frac{R_{\tr{cut}}}{\eta}.
    \end{equation}
    Numerical tests suggest that \cref{eq:fprefac_theoretical} works well for determining the cutoffs for calculating $[\mat{L}]_{abc}$, but leads to overestimation of the cutoffs for $[\mat{L}]_{ab}$.
    To that end, we drop the $R$ dependence and simply use
    \begin{equation}    \label{eq:fprefac_j2c}
        f_{2\tr{c}}(\eta, R_{\tr{cut}}, \Omega)
            = \eta^{-1}
    \end{equation}
    for solving \cref{eq:j2c_Rcut}.
    The numerical data justifying the choice of \cref{eq:fprefac_j2c} for the two-center integrals are shown in Fig.\ S3.

    \begin{figure}[t]
        \centering
        \includegraphics[width=0.7\linewidth]{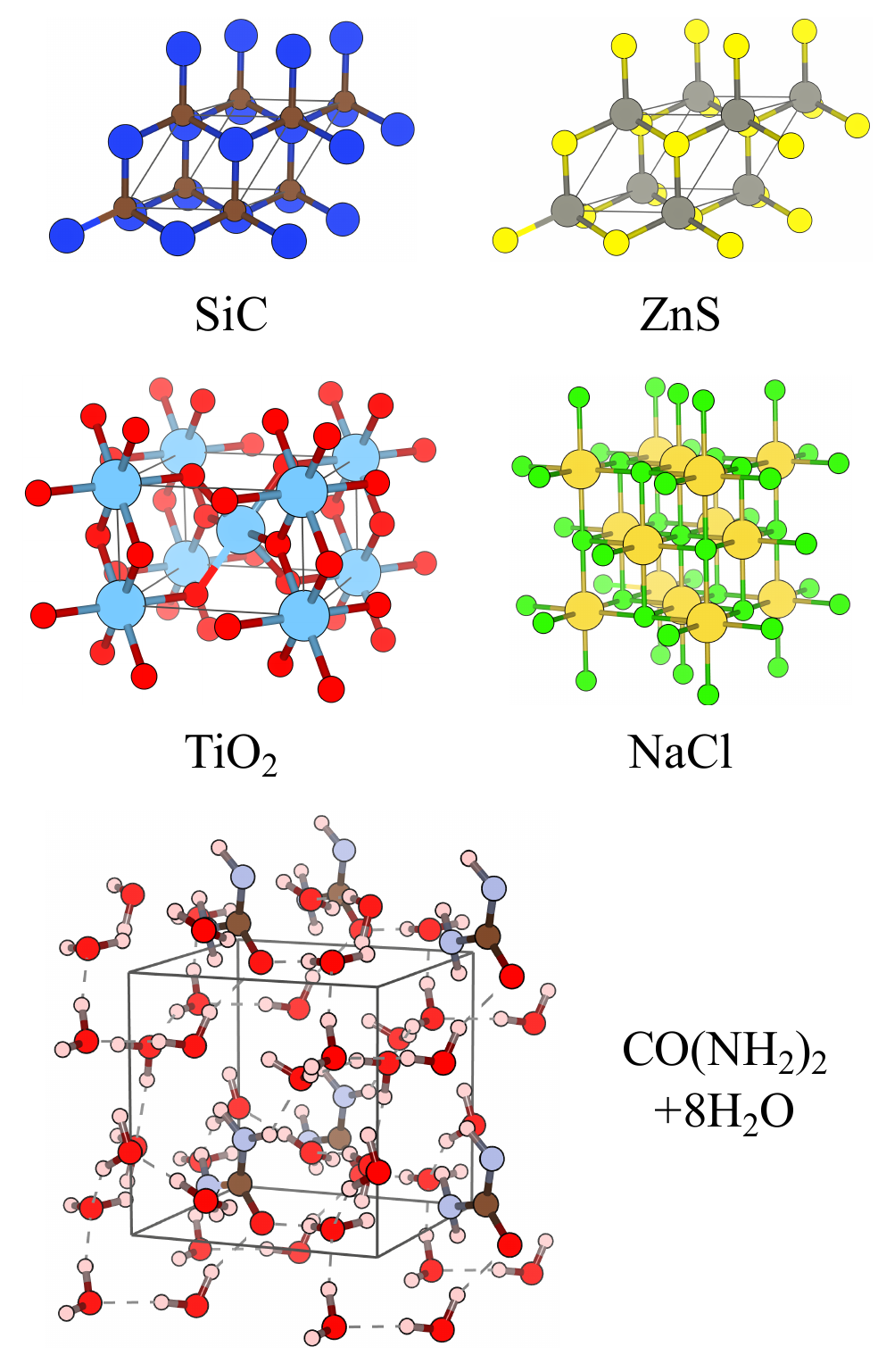}
        \caption{Structure of the systems tested in this work. The unit simulation cell is indicated by the grey box in each case.}
        \label{fig:lat}
    \end{figure}

\section{Computational details}
\label{sec:computational_details}

    We implemented all the estimators derived in \cref{subsec:j2cest,subsec:j3cest} as well as \cref{alg:j2c,alg:j3c} for calculating the periodic two-center and three-center SR ERIs in the PySCF software package \cite{Sun18WIRCMS}.
    We checked the correctness of our implementation by verifying that the calculated periodic integrals match those from the analytic Fourier transform (AFT) approach \cite{Sun17JCP} for small systems (limited by the high computational cost of AFT).
    The cutoff equations~(\ref{eq:j2c_Rcut}), (\ref{eq:j3c_dcut}), and (\ref{eq:j3c_Rcut}) are solved numerically using a binary search algorithm.
    The bin size $\Delta d$ for grouping the bra AO pairs is set to be $1$~\AA.
    All cutoffs and important cell (pair) data, including $\{R_{ab}^{\tr{cut}}\}$ and $\{\mc{M}_a\}$ for the two-center integrals (\cref{subsubsec:pbcj2c}) and $\{d_{ab}^{\tr{cut}}\}$, $\{R_{abc}^{\tr{cut}}(d_{ab}^{(i)})\}$,
    and $\{\mc{S}_{ab}^{(i)}\}$ for the three-center integrals (\cref{subsubsec:pbcj3c}), are precomputed and kept in memory before performing the lattice sum by \cref{alg:j2c,alg:j3c}.
    The cost of the precomputation is in general only a small fraction of the subsequent lattice sum (up to a few percent in the worst cases).

    \begin{table}[!b]
        \centering
        \caption{The exponent of the most diffuse orbital of given angular momentum for all basis sets used in this work. For each element, the first line is the AO basis, and the second line is the corresponding auxiliary basis ("JK" for the JKFIT basis and "ET" for the even tempered basis).}
        \label{tab:expn_AO}
        \begin{tabular}{lcccccccc}
            \hline\hline
            \multirow{2}*{Element} &
                    \multirow{2}*{Basis} & \multicolumn{7}{c}{$l$} \\
            \cline{3-9}
            & & $0$ & $1$ & $2$ & $3$ & $4$ & $5$ & $6$ \\
            \hline
            H & DZ & $0.122$ & $0.727$ & & & & & \\
              & JK & $0.284$ & $0.502$ & $0.713$ & & & & \\
            C & DZ & $0.160$ & $0.152$ & $0.550$ & & & & \\
              & JK & $0.192$ & $0.203$ & $0.200$ & $0.416$ & & & \\
            N & DZ & $0.225$ & $0.218$ & $0.817$ & & & & \\
              & JK & $0.273$ & $0.282$ & $0.290$ & $0.609$ & & & \\
            O & DZ & $0.302$ & $0.275$ & $1.185$ & & & & \\
              & JK & $0.339$ & $0.367$ & $0.356$ & $0.825$ & & & \\
            Na & DZ & $0.023$ & $0.021$ & $0.097$ & & & & \\
              & ET & $0.046$ & $0.044$ & $0.041$ & $0.089$ & $0.195$ & & \\
            Si & DZ & $0.092$ & $0.088$ & $0.275$ & & & & \\
              & JK & $0.132$ & $0.185$ & $0.170$ & $0.237$ & & & \\
            S & DZ & $0.157$ & $0.141$ & $0.479$ & & & & \\
              & JK & $0.216$ & $0.263$ & $0.242$ & $0.396$ & & & \\
            Cl & DZ & $0.194$ & $0.162$ & $0.600$ & & & & \\
              & JK & $0.234$ & $0.306$ & $0.275$ & $0.499$ & & & \\
            Ti & DZ & $0.025$ & $0.029$ & $0.052$ & $0.279$ & & & \\
              & ET & $0.051$ & $0.055$ & $0.059$ & $0.078$ & $0.104$ & $0.241$ & $0.558$ \\
            Zn & DZ & $0.038$ & $0.046$ & $0.252$ & $1.462$ & & & \\
              & ET & $0.075$ & $0.084$ & $0.093$ & $0.216$ & $0.503$ & $1.213$ & $2.923$ \\
            Zn & TZ & $0.038$ & $0.046$ & $0.252$ & $1.485$ & $4.114$ & & \\
              & ET & $0.075$ & $0.084$ & $0.093$ & $0.216$ & $0.503$ & $1.223$ & $2.970$ \\
            Zn & QZ & $0.037$ & $0.039$ & $0.159$ & $0.853$ & $2.069$ & $4.414$ & \\
              & ET & $0.074$ & $0.076$ & $0.078$ & $0.157$ & $0.319$ & $0.737$ & $1.706$ \\
             \hline
        \end{tabular}
    \end{table}

    We assess both the accuracy of the estimators and the computational cost of the lattice sum based on them over a test set of four three-dimensional solids and a water-solvated urea molecule as shown in \cref{fig:lat} (Cartesian coordinates in Supporting Information).
    The Dunning's cc-pVXZ basis set \cite{Dunning89JCP,Woon93JCP,Balabanov05JCP,Balabanov06JCP,Prascher11TCA} (abbreviated as "XZ" henceforth) is chosen as the AO basis.
    Specifically, we consider DZ for all systems and also TZ and QZ for ZnS.
    The cc-pVXZ-JKFIT basis set \cite{Weigend02PCCP} is used as the auxiliary basis for all elements except for Na, Ti, and Zn, for which the JKFIT basis set is not defined and we use the even tempered basis functions generated by PySCF with a progression factor $\beta = 2.0$ (details in Supporting Information).
    The exponents of the most diffuse orbitals of all basis sets used in this work are summarized in \cref{tab:expn_AO}.
    A series of $\omega$ values ranging from $0.1$ to $1$ are tested, which cover both the value used by the HSE functional ($\omega = 0.11$) and those commonly used by the range-separated Gaussian density fitting \cite{Ye21JCP} (RSGDF).
    These choices (atom types, crystal structures, basis sets, and $\omega$ values) together make the numerical study of this work cover a wide range of parameters.

    We measure the accuracy of the estimators in two ways.
    First, we calculate the intrinsic bra-ket cutoffs, $\{\tilde{R}_{ab}^{\tr{cut}}\}$ for the two-center case and $\{\tilde{R}_{abc}^{\tr{cut}}(d_{ab}^{(i)})\}$ for the three-center case, by solving \cref{eq:j2c_Rcut} and \cref{eq:j3c_Rcut} without the $f$-prefactor using both the exact SR ERIs and our estimators
    [the orbital orientation effect in the exact $\mc{J}_{abc}(\bm{R},\bm{d})$ is accounted for by averaging over three randomly generated configurations for each $(R,d)$].
    The error of the estimated cutoffs then reflects directly the accuracy and tightness of the corresponding estimators.
    Second, we calculate the maximum absolute error (MAE) of the periodic SR ERI tensor, $[\mat{L}]_{ab}$ and $[\mat{L}]_{abc}$, computed using \cref{alg:j2c,alg:j3c} based on our estimators.
    This provides an indirect but more practical measure of the accuracy of the estimators for their applications to periodic systems.

    Ideally, the error should be computed against the exact periodic integrals from the infinite lattice sum, which is unfortunately not possible in practice.
    To that end, we truncate the lattice sum using the most accurate estimators ($O_lv_l$ and ME for two- and three-center integrals, respectively, justified by the numerical data in \cref{sec:results_and_discussions}) with a tight target precision of $\epsilon = 10^{-12}$, and then perform the lattice sum without any screening, i.e.,\ with the ``if'' statements in \cref{alg:j2c,alg:j3c} always set to true.
    For three-center integrals, this essentially amounts to using only the rigorous upper bound provided by the Schwarz inequality to discard unimportant AO pairs.
    We verified that internal consistency is achieved in all cases: the MAEs calculated as above are essentially the same as those calculated against the integrals obtained with $\epsilon = 10^{-12}$ and with the screening based on the same estimator.

    The computational cost is measured by either the number of integrals being evaluated (i.e.,\ the number of times where the ``if'' statements in \cref{alg:j2c,alg:j3c} are evaluated to true) or the actual CPU time spent on the lattice sum.
    In this work, we will focus on the cost of the three-center integrals alone because the cost of evaluating the two-center integrals is essentially negligible in our applications.

\section{Results and discussions}
\label{sec:results_and_discussions}

    \begin{figure}[t]
        \centering
        \includegraphics[width=0.6\linewidth]{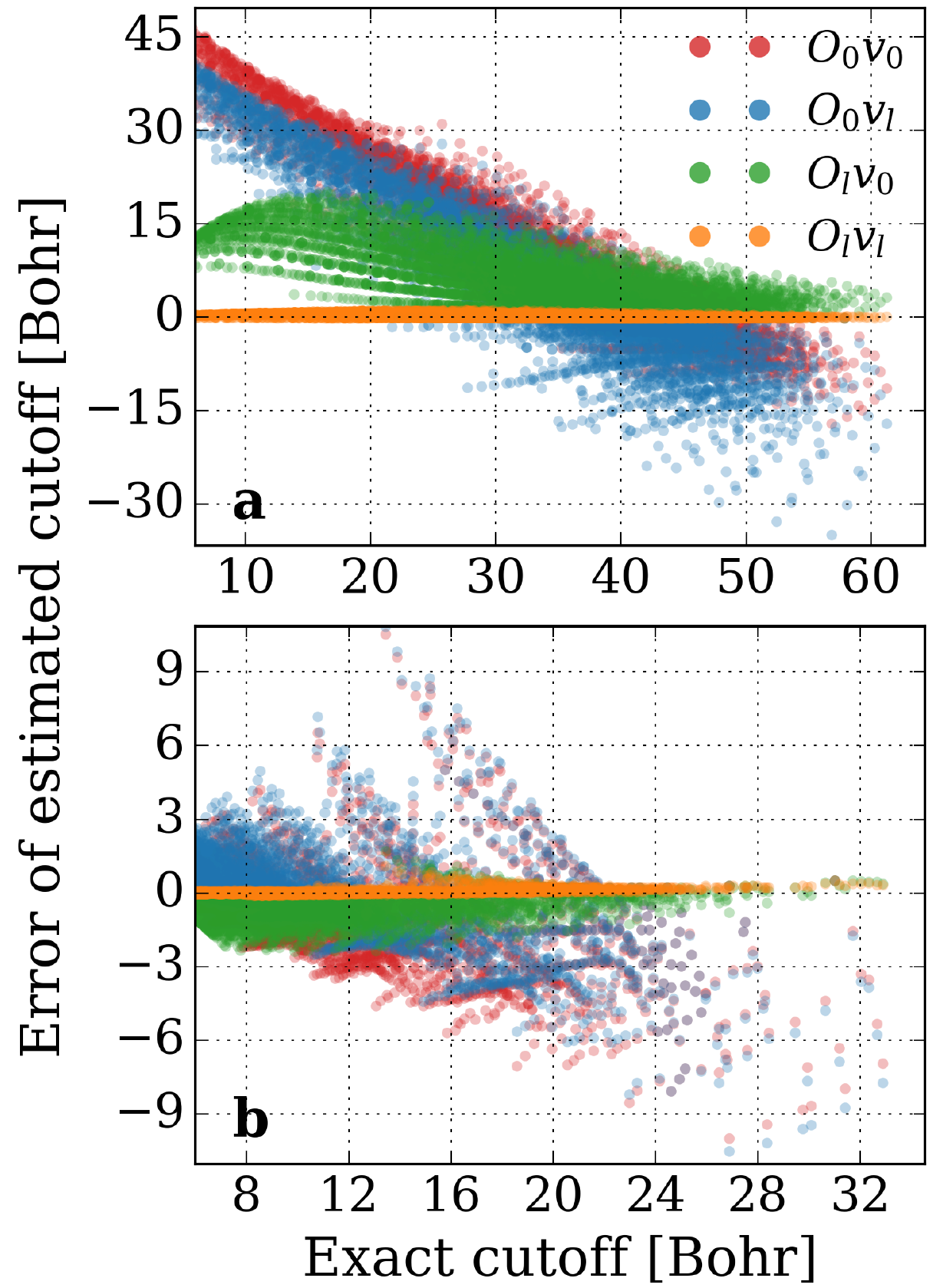}
        \caption{Error of the cutoffs for two-center SR ERIs estimated by the four estimators from \cref{subsec:j2cest} with $\epsilon = 10^{-10}$ plotted against the exact cutoffs for \ce{TiO2}/DZ and two values of $\omega$, $0.1$ (a) and $1$ (b).}
        \label{fig:j2c_cutoff}
    \end{figure}

    \begin{figure*}[t]
        \centering
        \includegraphics[width=0.8\linewidth]{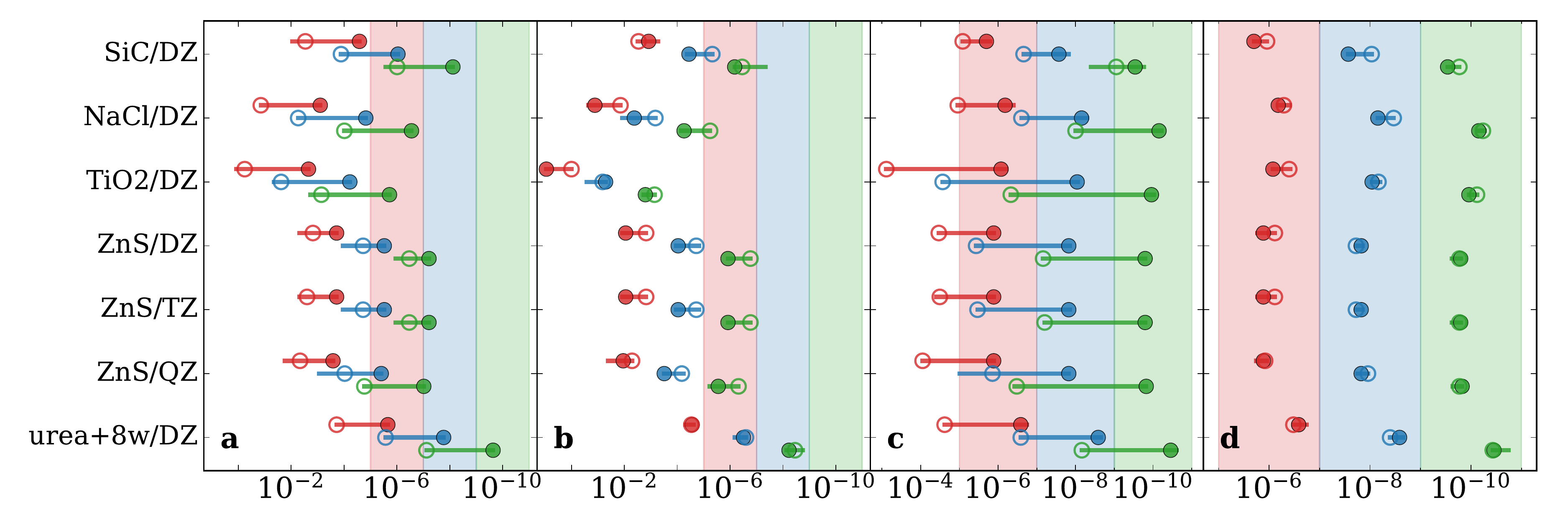}
        \caption{MAE of periodic two-center SR ERIs computed using \cref{alg:j2c} with four different two-center estimators for integral screening: $O_0 v_0$ (a), $O_0 v_l$ (b), $O_l v_0$ (c), and $O_l v_l$ (d). Different colors are for different target precision, $\epsilon = 10^{-6}$ (red), $10^{-8}$ (blue), and $10^{-10}$ (green), and the shaded area highlights $0.1\epsilon \sim 10\epsilon$ for the corresponding color.
        For each system and target precision, the horizontal bar shows the distribution of the MAEs for $\omega = 0.1, 0.2, \cdots, 1$, where $\omega = 0.1$ and $1$ are highlighted as a filled and hollow circles, respectively.}
        \label{fig:j2c_err}
    \end{figure*}

    \subsection{Accuracy of two-center estimators}
    \label{subsec:acc_j2cest}

    We first study the accuracy of the four two-center estimators from \cref{subsec:j2cest}.
    In \cref{fig:j2c_cutoff}, we show the error of the estimated cutoffs for \ce{TiO2}/DZ with $\epsilon = 10^{-10}$ for $\omega=0.1$ (a) and $\omega=1$ (b);
    the same plots for other systems are shown in Fig.\ S4.
    An immediate conclusion one can draw from these plots is that the $O_l v_l$ estimator~(\ref{eq:j2cest_Olvl}) is very tight, predicting essentially the exact cutoffs in all cases.
    As a result, the periodic integrals $[\mat{L}]_{ab}$ computed based on the $O_l v_l$ estimator display highly controlled accuracy for all systems and all values of $\omega \in [0.1,1]$, as shown in \cref{fig:j2c_err}(d).
    The high accuracy of the $O_l v_l$ estimator is due to the correct treatment of the $l$-dependence in both the orbital multipoles and the effective potential: ignoring the $l$-dependence in either or both cases generally leads to higher errors as can be seen in \cref{fig:j2c_cutoff} (red, blue and green dots) and \cref{fig:j2c_err}(a)--(c).
    These results can be understood as follows.

    The $l$-dependence in the orbital multipoles depends strongly on the orbital exponents as $\zeta^{-(l+3/2)}$ [\cref{eq:Oal}].
    For diffuse orbitals ($\zeta \ll 1$) of high angular momentum, approximating $O_l$ by $O_0$ leads to significant underestimation. This explains the negative errors of the $O_0 v_0$ (red) and the $O_0 v_l$ (blue) estimators in \cref{fig:j2c_cutoff} when the exact cutoffs are large, which produce large errors in the corresponding periodic integrals in \cref{fig:j2c_err}(a) and (b).
    Also, the highest error of the periodic integrals is seen in \ce{TiO2} in both cases, because the even tempered auxiliary basis for Ti/DZ has very diffuse shells ($\zeta \lessapprox 0.1$) up to $l = 4$ (\cref{tab:expn_AO}).
    On the other hand, for non-$s$-type compact orbitals ($\zeta > 1$), the $O_0$ approximation overestimates the true integrals as is clear from the large positive cutoff errors of the two $O_0$ estimators (red and blue) in \cref{fig:j2c_cutoff} when the exact cutoffs are small.
    This does no harm to the accuracy of the periodic integrals over these orbitals but increases the computational cost.

    The $l$-dependence in the effective potential shows a strong $\omega$-dependence.
    For $\omega$ small, the SR Coulomb potential resembles the full Coulomb potential, and approximating $v_l$ by $v_0$ overestimates the integrals as in the classical multipole interaction and leads to positive errors for the $O_l v_0$ estimator (green) in \cref{fig:j2c_cutoff}(a).
    This also explains two trends observed for the filled circles ($\omega = 0.1$) in \cref{fig:j2c_err}: (i) $O_0 v_0$ is more accurate than $O_0 v_l$ [\cref{fig:j2c_err}(a) and (b)], and (ii) $O_l v_0$ is as accurate as $O_l v_l$ [\cref{fig:j2c_err}(c) and (d)].
    As $\omega$ increases, however, the $l$-dependence in $v_l$ starts to deviate from that of the classical multipole interaction and the $v_0$ approximation tends to underestimate the integrals.
    Consequently, negative errors are seen for the cutoffs of the $O_l v_0$ estimator (green) in \cref{fig:j2c_cutoff}(b) and quick growth of the MAEs of $[\mat{L}]_{ab}$ with $\omega$ is observed for the $O_0 v_0$ and the $O_l v_0$ estimators in \cref{fig:j2c_err}(a) and (c) (hollow circles), respectively.

    To summarize the results for two-center estimators, the numerical data presented in \cref{fig:j2c_cutoff,fig:j2c_err} confirm the tightness of the $O_l v_l$ estimator and the accuracy of the resulting periodic integrals, hence justifying its use in \cref{subsec:j3cest} for obtaining the three-center estimators, whose performance is discussed in the next section.

    \begin{figure*}[t]
        \centering
        \includegraphics[width=0.8\linewidth]{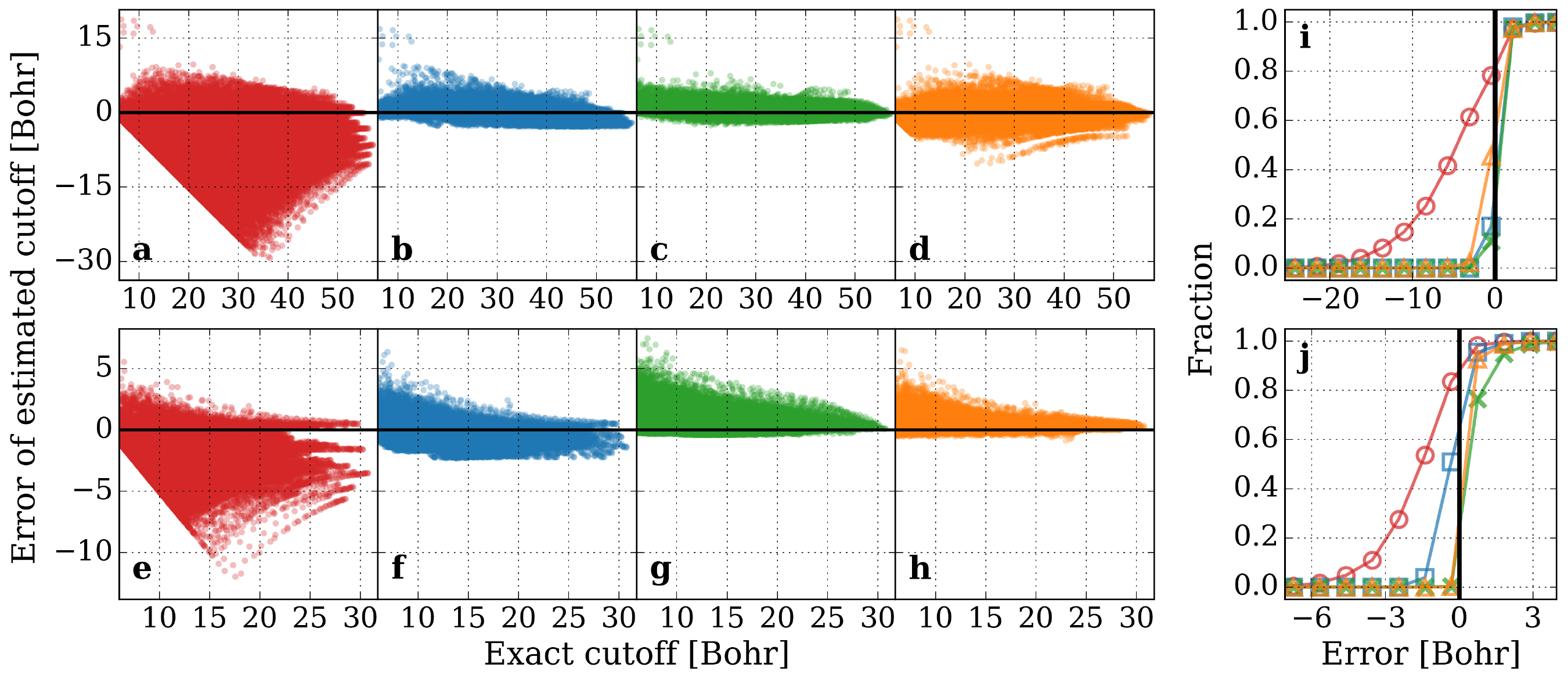}
        \caption{Error of the cutoffs for three-center SR ERIs estimated by the four estimators from \cref{subsec:j3cest}, ISF (a,e), ISF$Q_0$ (b,f), ISF$Q_l$ (c,g), and ME (d,h), with $\epsilon = 10^{-10}$ plotted against the exact cutoffs for \ce{TiO2}/DZ and two values of $\omega$, $0.1$ (a -- d) and $1$ (e -- h).
        Panels (i) and (j) show the fraction of estimated cutoffs with error lower than the abscissa.}
        \label{fig:j3c_cutoff}
    \end{figure*}

    \begin{figure*}[t]
        \centering
        \includegraphics[width=0.8\linewidth]{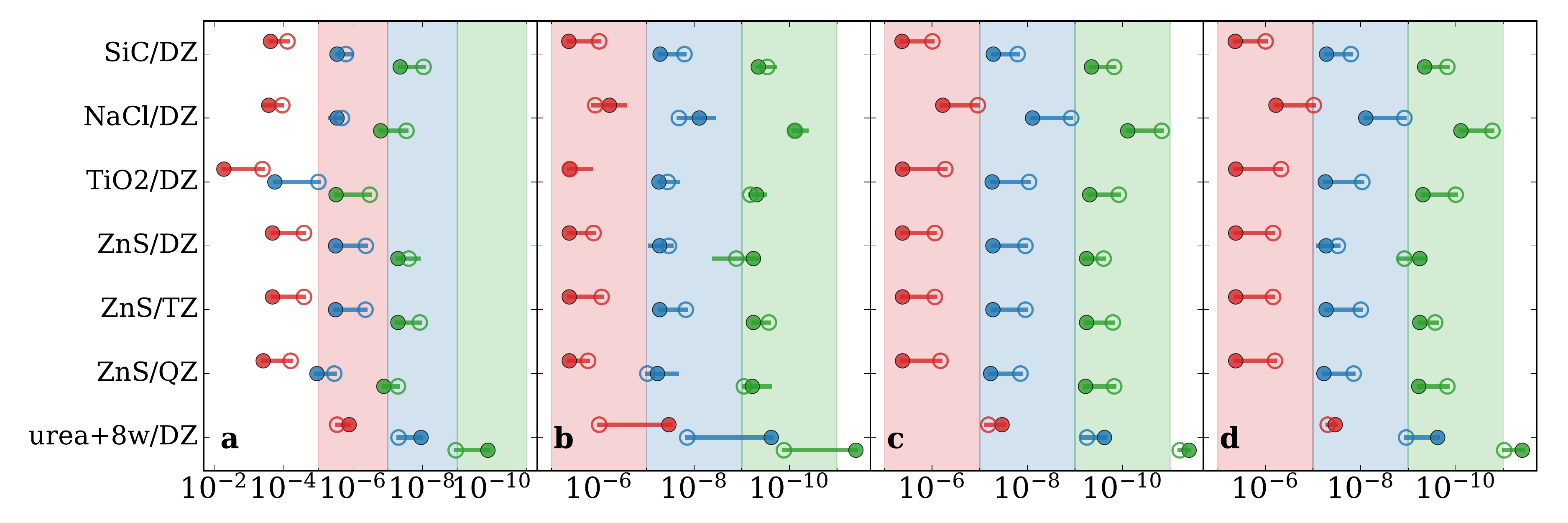}
        \caption{Same plot as \cref{fig:j2c_err} for the MAE of periodic three-center SR ERIs computed using \cref{alg:j3c} with different three-center estimators for integral screening: ISF (a), ISF$Q_0$ (b), ISF$Q_l$ (c), and ME (d).}
        \label{fig:j3c_err}
    \end{figure*}

    \subsection{Accuracy of three-center estimators}
    \label{subsec:acc_j3cest}

    In \cref{fig:j3c_cutoff}, we show the error of the cutoffs computed by the four three-center estimators from \cref{subsec:j3cest} for \ce{TiO2}/DZ with $\epsilon = 10^{-10}$ and $\omega = 0.1$ and $1$;
    the same plots for other systems can be found in Figs.\ S5 and S6.
    An immediate conclusion from these plots is that the ISF estimator~(\ref{eq:j3cest_ISF}) significantly underestimates the true integrals in all cases.
    For \ce{TiO2}, it underestimates about 80\% of the cutoffs as shown in \cref{fig:j3c_cutoff}(i) and (j) (red circles).
    The underestimation by the ISF estimator comes from ignoring the higher-order multipoles of the bra AO product and
    results for \ce{TiO2} clearly show more severe underestimation for largelr $l_{ab}$ (Fig.\ S7).
    As a result, the MAEs of the periodic three-center SR ERIs computed based on the ISF estimator are two to four order of magnitude higher than the target precision as shown in \cref{fig:j3c_err}(a).
    One exception is the solvated urea molecule, where the low packing density makes the $f$-prefactor~(\ref{eq:fprefac_theoretical}) overestimate the truncation error, which cancels the underestimation by the ISF estimator and leads to MAEs $\lessapprox 10\epsilon$ in this case.
    In general, one should not rely on such fortuitous error cancellation.

    The underestimation by the ISF estimator is largely corrected by the three other estimators that account for the higher-order multipoles of the bra AO product, as can be seen in \cref{fig:j3c_cutoff}(b)--(d) and (f)--(h).
    As expected from the discussion in \cref{subsubsec:ISFQl}, the ISF$Q_l$ estimator~(\ref{eq:j3cest_ISFQl}) reduces essentially to the ISF$Q_0$ estimator~(\ref{eq:j3cest_ISFQ0}) for small $\omega$ [\cref{fig:j3c_cutoff}(b) and (c)],
    but corrects the slight underestimation of the latter for large $\omega$ [\cref{fig:j3c_cutoff}(f) and (g)].
    The ME estimator~(\ref{eq:j3cest_ME}), which includes all terms in the multipole expansion with approximate GPT coefficients~(\ref{eq:Talmi_coeff_zz})~and~(\ref{eq:Talmi_coeff_d0}), is the most accurate among the three for large $\omega$ [\cref{fig:j3c_cutoff}(h)], but shows slight underestimation for small $\omega$ [\cref{fig:j3c_cutoff}(d)].
    Overall, the three estimators have similar performance, and the difference between them is at most modest.
    This is also reflected by the high accuracy of the periodic integrals computed based on these estimators as shown in \cref{fig:j3c_err}(b)--(d), where the MAEs typically fall in the range of $0.1\epsilon \sim 10\epsilon$ (with the solvated urea the only exception where the MAE is lower than $\epsilon$ for the reason discussed above).

    In summary, the data in \cref{fig:j3c_cutoff,fig:j3c_err} suggest that ISF$Q_0$, ISF$Q_l$, and ME are all accurate estimators for three-center SR ERIs.
    In the next section, we will assess them based on the computational cost of the screened lattice sum.

    \subsection{Computational cost of the lattice sum for three-center integrals}
    \label{subsec:computational_cost}

    \begin{figure*}[t]
        \centering
        \includegraphics[width=0.9\linewidth]{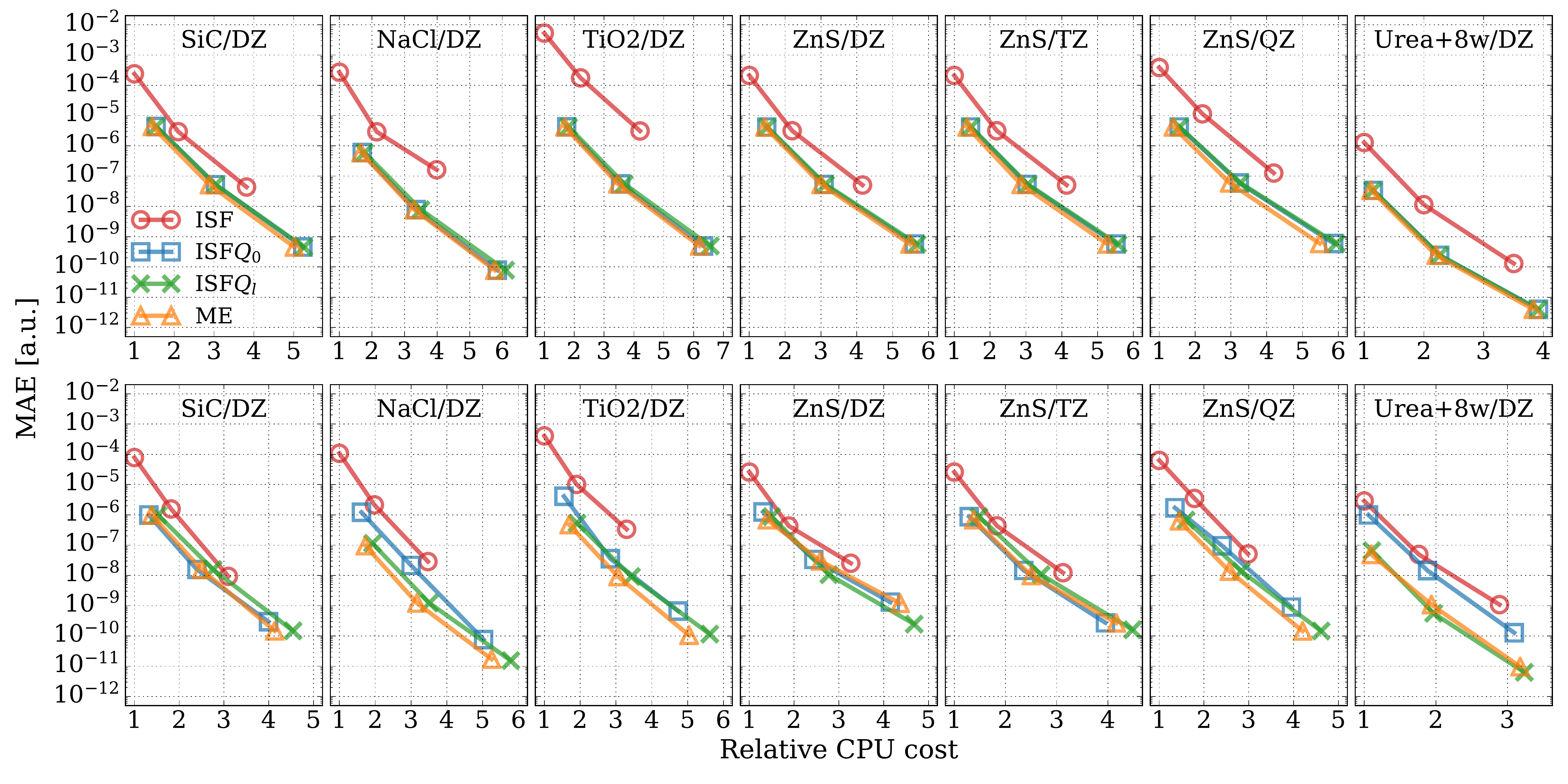}
        \caption{The MAE data taken from \cref{fig:j3c_err} plotted as a function of the CPU cost measured by the total number of integrals being calculated in the lattice sum (relative to the cost of ISF/$\epsilon = 10^{-6}$) for $\omega = 0.1$ (upper panels) and $1$ (lower panels).}
        \label{fig:j3c_prec_vs_nshlpr}
    \end{figure*}

    \subsubsection{Computational efficiency}

    The results of the previous section clearly show increasing accuracy of the three-center estimator as its complexity grows.
    However, this alone does not justify the practical use of the more accurate estimators (ISF$Q_0$, ISF$Q_l$, and ME) because, although tedious, one can always empirically adjust the $\epsilon$ parameter for an inaccurate estimator (ISF in this case) to achieve the same accuracy for a specific system.
    We thus need to compare the computational cost of the lattice sum based on different estimators \emph{for achieving the same accuracy}.
    We measure the computational cost by counting the number of integrals being evaluated in the lattice sum for computing the entire $[\mat{L}]_{abc}$ tensor and study its relation to the MAE data presented in \cref{fig:j3c_err}.

    The results are plotted in \cref{fig:j3c_prec_vs_nshlpr} for the two extreme cases, $\omega = 0.1$ and $1$.
    In both cases, the ISF curves of all systems lie to the upper right of the curves of the other three estimators, suggesting consistently higher computational cost of the lattice sum based on the ISF estimator for achieving the same accuracy as the other three.
    For small $\omega$ (upper panels in \cref{fig:j3c_prec_vs_nshlpr}), the saving in computational cost by using the more accurate estimators is as large as a factor of two in the most challenging case (\ce{TiO2}).
    For large $\omega$ (lower panels in \cref{fig:j3c_prec_vs_nshlpr}), however, only the ME estimator maintains the large saving against ISF for all systems, while the ISF$Q_0$ and ISF$Q_l$ curves both move closer to the ISF curve in some cases, indicating a loss of computational efficiency of the two Schwarz-$Q$ integral-based estimators for large $\omega$.
    This observation confirms that the ME estimator is tighter than the ISF$Q_0$ and ISF$Q_l$ estimators when $\omega$ is large.

    \subsubsection{Scaling with $\omega$}

    \begin{figure}[b]
        \centering
        \includegraphics[width=0.6\linewidth]{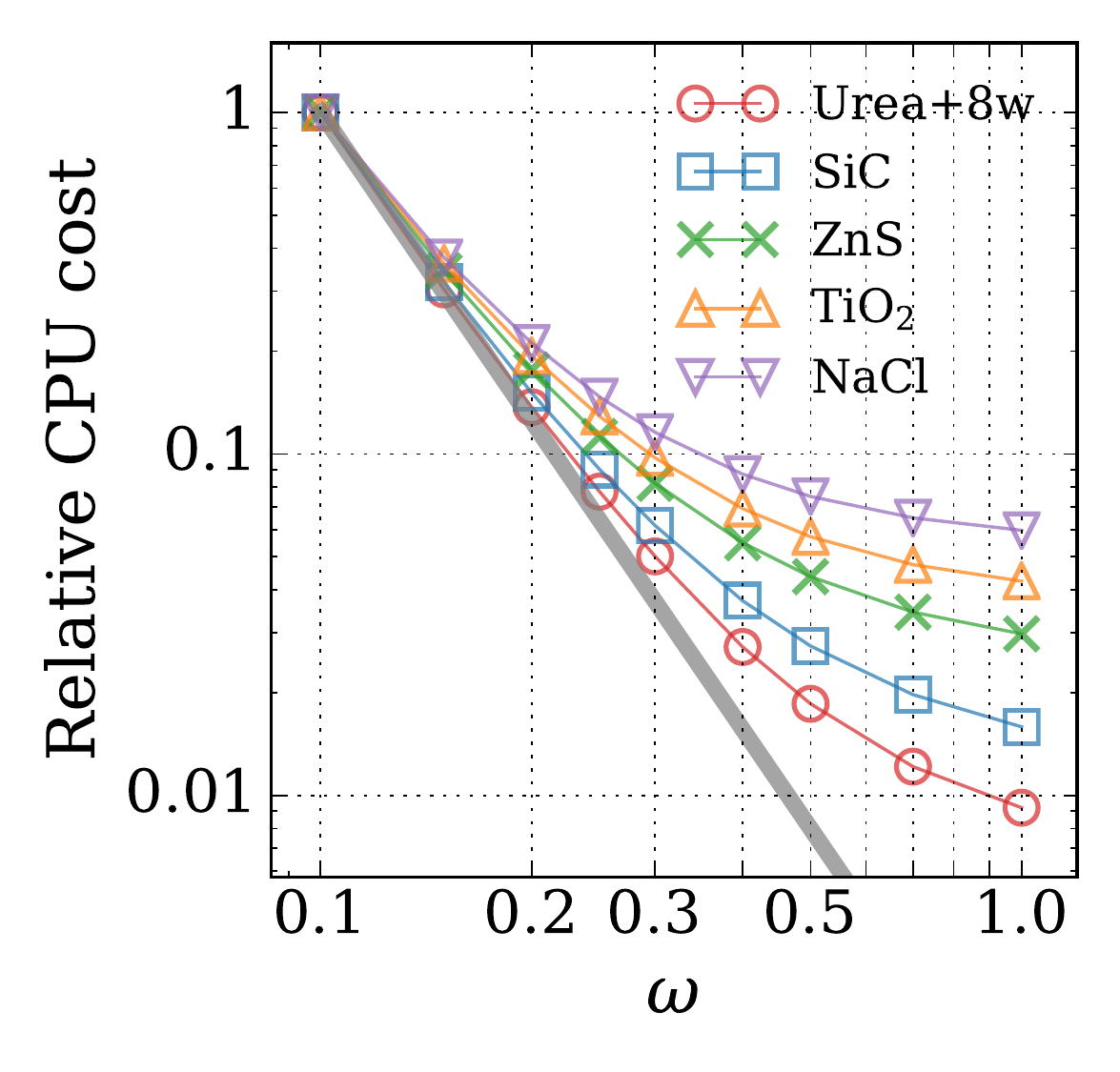}
        \caption{Relative CPU cost of the lattice sum for computing $[\mat{L}]_{abc}$ based on the ME estimator [\cref{eq:j3cest_ME}] with $\epsilon = 10^{-8}$ plotted as a function of $\omega$ (in log-log scale) for the five test systems in the DZ basis.
        The grey line indicates a perfect $\omega^{-3}$ scaling of the relative CPU cost.}
        \label{fig:j3c_nshlpr_vs_omega}
    \end{figure}

    We next investigate how the cost of the screened lattice sum scales with $\omega$.
    Given that the characteristic decay length of the SR Coulomb potential is roughly $\omega^{-1}$, one may expect $\omega^{-3}$ scaling for three-dimensional systems.
    However, as discussed in \cref{subsec:j2cest}, the decay of the final SR ERIs is described instead by an effective potential, $v_l(\eta, R)$ [\cref{eq:vl_def}], where the bare $\omega$ is replaced, for three-center integrals, by $\eta_{abc\omega}$ [\cref{eq:etaabcw_def}],
    which depends not only on $\omega$ but also on the exponents of the orbitals.
    This means that different elements in the $[\mat{L}]_{abc}$ tensor need different computational effort, and the term with the smallest $\eta_{abc\omega}$ represents the computational bottleneck.
    The smallest $\eta_{abc\omega}$ (call it $\eta_{abc\omega}^{\tr{min}}$) for a given system is roughly the minimum of $\omega^2$ and half the smallest AO exponent, $\eta_{\tr{AO}}^{\tr{min}} \equiv \zeta_{\tr{AO}}^{\tr{min}}/2$.

    We predict the following two limits for the scaling of the computational cost with $\omega$.
    In the small $\omega$ regime where $\omega \ll [\eta_{\tr{AO}}^{\tr{min}}]^{1/2}$, we have $\eta_{abc\omega}^{\tr{min}} \approx \omega^2$ and we expect the ideal $\omega^{-3}$ scaling.
    In the large $\omega$ regime where $\omega \gg [\eta_{\tr{AO}}^{\tr{min}}]^{1/2}$, we have $\eta_{abc\omega}^{\tr{min}} \approx \eta_{\tr{AO}}^{\tr{min}}$ and we expect a plateau in the cost as a function of $\omega$.
    These asymptotic predictions are confirmed numerically in \cref{fig:j3c_nshlpr_vs_omega} for the ME estimator with $\epsilon =10^{-8}$;
    other choices of the estimator and $\epsilon$ lead to essentially identical plots (Fig.\ S8).
    In all cases, the cost follows the ideal $\omega^{-3}$ decay (grey line) but then begins to plateau as $\omega$ increases.
    The crossover point between these asymptotic behaviors (and therefore also the plateau height) are consistent with the values of $[\eta_{\tr{AO}}^{\tr{min}}]^{1/2}$ for these systems:
    $0.25$ for the solvated urea, $0.21$ for SiC, $0.14$ for ZnS, $0.11$ for \ce{TiO2}, and $0.10$ for NaCl (all calculated using the data from \cref{tab:expn_AO}).

    \subsubsection{Scaling with system size}

    \begin{figure}[b]
        \centering
        \includegraphics[width=0.6\linewidth]{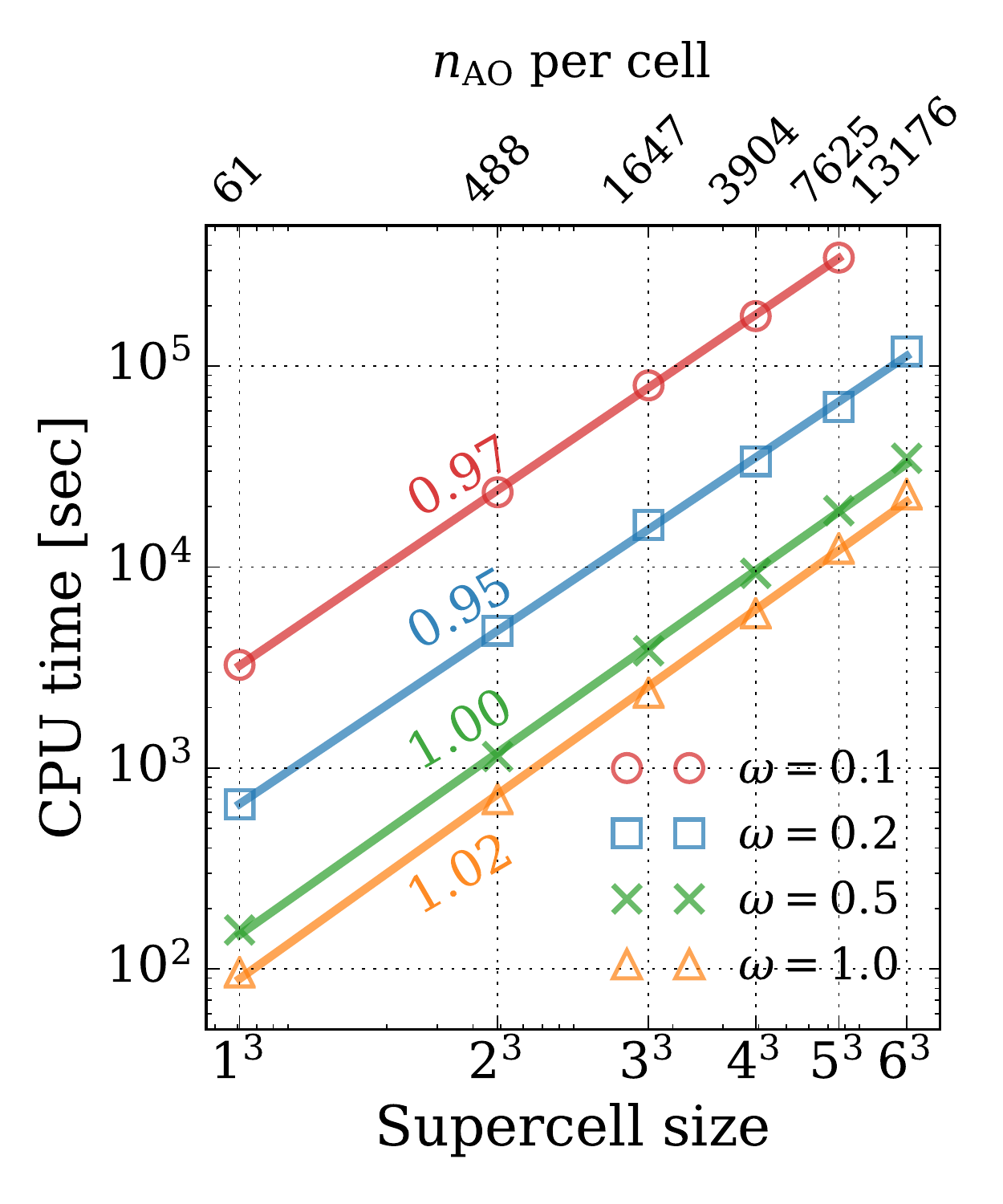}
        \caption{CPU time for the lattice sum for computing $[\mat{L}]_{abc}$ using the ME estimator with $\epsilon=10^{-8}$ for ZnS/DZ of different supercell size and different values of $\omega$.
        Each line is a linear fit for the data of the same color.
        The apparent scaling factors obtained from the fitting are listed aside, all showing linear scaling with system size.
        The number of AOs per supercell is also shown in the top abscissa.}
        \label{fig:j3c_tcpu_zns}
    \end{figure}

    In periodic calculations, an infinite system is approximated by a finite simulation supercell subject to Born-von Karman periodic boundary conditions (\cref{subsec:pbcj23c}).
    This introduces a finite-size error that must be removed by increasing the size of the simulation supercell (or equivalently the density of Brillouin zone sampling). \cite{Gygi86PRB,Paier06JCP,Spencer08PRB,Broqvist09PRB,Guidon09JCTC,Sundararaman13PRB,Azadi15JCP,McClain17JCTC}
    Therefore, in this final section, we study how the cost of the screened lattice sum scales with the size of the supercell.

    For each element of the $[\mat{L}]_{abc}$ tensor, the number of integrals that pass the screening is $\sum_{i}^{n_{d_{ab}}} |\mc{S}_{abc}^{(i)}|$, where $\mc{S}_{abc}^{(i)}$ is defined in \cref{eq:Sabc_i}.
    Thus, the total computational cost for executing \cref{alg:j3c} is proportional to
    \begin{equation}    \label{eq:numsigint}
        N_{\tr{I}}
            = \sum_{a,b}^{n_{\tr{AO}}} \sum_{i}^{n_{d_{ab}}}
            \sum_{c}^{n_{\tr{aux}}} |\mc{S}_{abc}^{(i)}|,
    \end{equation}
    where both $n_{\tr{AO}}$ and $n_{\tr{aux}}$ grow linearly with the system size while $\{n_{d_{ab}}\}$ remain constant, which seems to suggest cubic scaling with system size.
    However, since the cutoffs, $\{d_{ab}^{\tr{cut}}\}$ [\cref{eq:j3c_dcut}] and $\{R_{abc}^{\tr{cut}}\}$ [\cref{eq:j3c_Rcut}], depend only on the nature of the orbitals and are independent of the system size,
    we expect a quadratic decrease of the number of elements in each $\mc{S}_{abc}^{(i)}$ with system size according to \cref{eq:Sabc_i}.
    This suggests that the overall scaling with system size is linear if integral screening is performed as described in \cref{subsubsec:pbcj3c}.

    To confirm this analysis, we perform a series of supercell calculations for ZnS/DZ.
    A supercell of size $n_{\tr{p}}^3$ is constructed by repeating a primitive cell $n_{\tr{p}}$ times in each of the three dimensions.
    The CPU time of the lattice sum for computing the entire $[\mat{L}]_{abc}$ tensor using the ME estimator with $\epsilon=10^{-8}$ is shown in \cref{fig:j3c_tcpu_zns} for $n_{\tr{p}} = 1-6$ and four different values of $\omega$.
    Results generated using the ISF$Q_0$ and the ISF$Q_l$ estimators show similar trends (Fig.\ S9).
    For all choices of $\omega$, the results demonstrate nearly perfect linear scaling with the supercell size.

\section{Concluding remarks}
\label{sec:concluding_remarks}

    In summary, we derived distance-dependent estimators for the two-center and three-center SR ERIs over atom-centered GTOs.
    Performance was assessed by the accuracy of the periodic two-center and three-center SR ERIs, which are calculated using the estimators to screen the integrals appearing in the lattice summation.
    Based on the numerical data collected for systems that cover a wide range of parameters including the atom types, crystal structures, orbital exponents and angular momenta, and the range-separation parameter $\omega$,
    we recommend the $O_l v_l$ estimator~(\ref{eq:j2cest_Olvl}) and the ME estimator~(\ref{eq:j3cest_ME}) for two-center and three-center SR ERIs, respectively.
    In the case of small $\omega$ (such as that used in the HSE functional), the ISF$Q_0$ estimator~(\ref{eq:j3cest_ISFQ0}) is also a good choice for three-center SR ERIs.
    We discussed why the computational scaling of the lattice sum for three-center integrals deviates from $\omega^{-3}$, and show that the cost scales linearly with system size for all tested values of $\omega$.

    Although we chose to demonstrate their use for periodic systems, the estimators derived in this work also should be useful in large molecular applications.
    We also expect our results to be useful for semiempirical methods, where two-center and three-center ERIs are typical, and phenomenologically screened or otherwise SR Coulomb interactions are commonly used.
    For \textit{ab initio} periodic calculations using our recently developed range-separated Gaussian density fitting \cite{Ye21JCP} (RSGDF), building the three-center SR ERI tensor represents one of the main computational bottlenecks--at least for Hartree-Fock and lower-order perturbation theory.
    We thus anticipate the estimators together with the algorithms for screening the lattice sum to improve the computational efficiency of such calculations.

\section*{Supplementary material}

See the supplementary material for (i) the performance of ME estimator~(\ref{eq:j3cest_ME}) for three-center SR ERIs with concentric bra pairs (i.e.,\ $d_{ab} = 0$), (ii) the effect of using different $f$-prefactors to solve \cref{eq:j2c_Rcut} on the performance of the $O_l v_l$ estimator, (iii) the error of the estimated two-center cutoffs, (iv) the error of the estimated three-center cutoffs, (v) AO angular momentum-resolved error plot of the estiamted three-center cutoffs, (vi) the CPU cost of the lattice sum for computing $[\mat{L}]_{abc}$ plotted as a function of $\omega$ for different choices of estimator and $\epsilon$, (vii) the CPU time of the lattice sum for computing $[\mat{L}]_{abc}$ plotted as a function of the supercell size of ZnS/DZ for the ISF$Q_0$ and the ISF$Q_l$ estimators, (viii) Cartesian coordinates of the test systems shown in \cref{fig:lat}, (ix) details of the even tempered basis functions for Na, Ti, and Zn.

\appendix
\section{Fourier transform of primitive GTOs and SR Coulomb potentials}
\label{app:fourier}

\begin{equation}
\begin{split}
    \tphi_{alm}(\bm{q})
        &= \int\md\bm{r}\,
        \me^{-\mi \bm{q}\cdot\bm{r}} \phi_{alm}(\bm{r})   \\
        &= \mi^{l} (\pi/\zeta_a)^{3/2} (2\zeta_a)^{-l} q^{l}
        \me^{-q^2/(4\zeta_a)} y_{l m}(\hat{\bm{q}})
\end{split}
\end{equation}
\begin{equation}
    \tilde{g}_{\omega}(\bm{q})
        = \int\md\bm{r}\,
        \me^{-\mi \bm{q}\cdot\bm{r}} g_{\omega}(\bm{r})
        = \frac{4\pi}{q^2} \big(
            1 - \me^{-q^2/(4\omega^2)}
        \big)
\end{equation}

\section{Asymptotic analysis of \cref{eq:I_def}}
\label{app:asymptotic}

    The integral~(\ref{eq:I_def}) can be analytically performed,
    \begin{equation}    \label{eq:I_res_Mfunc}
        I(R; L, l, \eta)
            = 2^{L-1} \pi^{1/2} \frac{\Gamma(\alpha)}{\Gamma(\beta)}
            \eta^{\alpha} R^{\beta-3/2}
            M(\alpha,\beta;-\eta R^2)
    \end{equation}
    where $\alpha = (L+l+1)/2$, $\beta = l + 3/2$, and $M(a,b;z)$ is the confluent hypergeometric function of the first kind.
    Using the asymptotic behavior of $M(a,b;z)$ at large $|z|$,
    \begin{equation}
        M(a,b;z)
            \sim \Gamma(b) \bigg[
                \frac{\me^{z} z^{a-b}}{\Gamma(a)} +
                \frac{(-z)^{-a}}{\Gamma(b-a)}
            \bigg],
    \end{equation}
    we obtain an asymptotic expression for \cref{eq:I_res_Mfunc} at large $R$,
    \begin{equation}    \label{eq:I_asymptotic}
        I(R; L, l, \eta)
            \sim 2^{L-1} \pi^{1/2} \bigg[
                \frac{\Gamma(\alpha)}{\Gamma(\beta-\alpha) R^{L+1}} -
                R^{L-2} \eta^{L-1/2} \me^{-\eta R^2}
            \bigg].
    \end{equation}
    The first term decaying as $R^{L+1}$ is the \emph{long-range} classical multipole interaction.
    As this term is $\eta$-independent, its contribution to \cref{eq:j2c_qint} vanishes when taking the difference.
    The second term decaying exponentially with $R$ is the \emph{short-range} interaction via orbital overlap.
    Its contribution to \cref{eq:j2c_qint} is non-vanishing but $l$-independent.

\section*{Acknowledgements}

HY thanks Dr.\ Qiming Sun for helpful discussions.
This work was supported by the National Science Foundation under Grant No.\
OAC-1931321.  We acknowledge computing resources from Columbia University's
Shared Research Computing Facility project, which is supported by NIH Research
Facility Improvement Grant 1G20RR030893-01, and associated funds from the New
York State Empire State Development, Division of Science Technology and
Innovation (NYSTAR) Contract C090171, both awarded April 15, 2010. The Flatiron
Institute is a division of the Simons Foundation.

\section*{Data availability statement}
The data that support the findings of this study are available from the
corresponding author upon reasonable request.

\bibliography{refs}

\raggedbottom

\end{document}


\maketitle

    \tableofcontents

    \vspace{3em}

    \hspace{2em}Note: figures and equations appearing in the main text will be referred as
``Fig.\ Mxxx'' and ``Eq.\ Mxxx'' in this Supplementary Material document.


    \section{Supplementary figures}

    \begin{figure}[!h]
        \centering
        \includegraphics[width=0.84\linewidth]{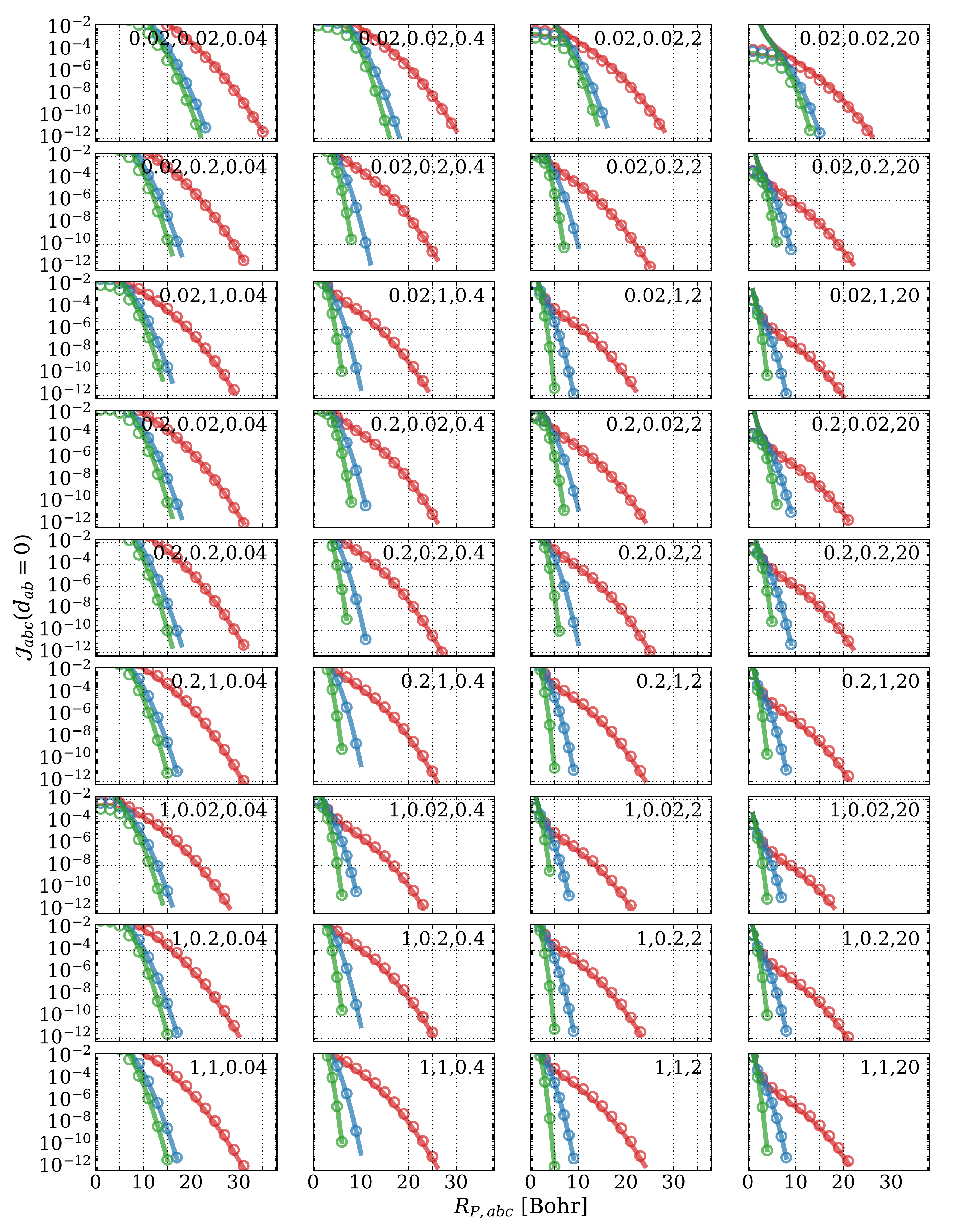}
        \caption{Frobenius norm of the three-center SR ERIs over all pGTOs [Eq.\ M27] for concentric bra pairs (i.e.,\ $d_{ab} = 0$) evaluated exactly (circles) and approximately by the ME estimator [Eq.\ M42] with the GPT coefficients determined using Eqs.\ M43 and M44.
        Three colors correspond to three representative values of $\omega$, $0.1$ (red), $0.3$ (blue), and $0.7$ (green).
        Each panel corresponds to a different choice of Gaussian exponents, $(\zeta_a,\zeta_b,\zeta_c)$, as specified by the three numbers at the upper right corner.
        $l_a = 1$, $l_b = 2$, and $l_c = 2$ are used.}
        \label{fig:j3cd0_l1_l2_laux2}
    \end{figure}

    \begin{figure}[!h]
        \centering
        \includegraphics[width=0.84\linewidth]{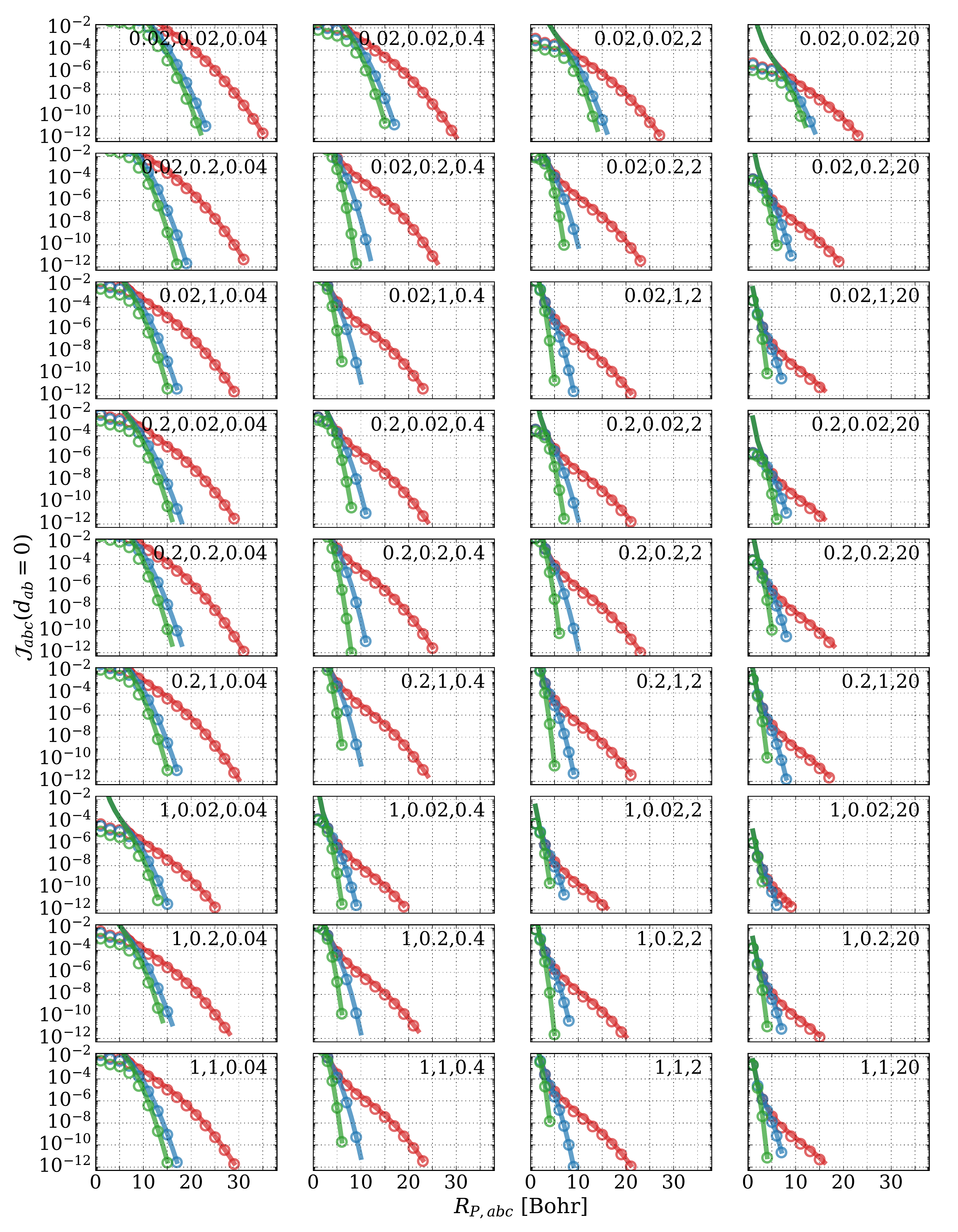}
        \caption{Same plot as \cref{fig:j3cd0_l1_l2_laux2} except for $l_a = 0$, $l_b = 3$, and $l_c = 3$.}
        \label{fig:j3cd0_l0_l3_laux3}
    \end{figure}

    \begin{figure}[!h]
        \centering
        \includegraphics[width=0.84\linewidth]{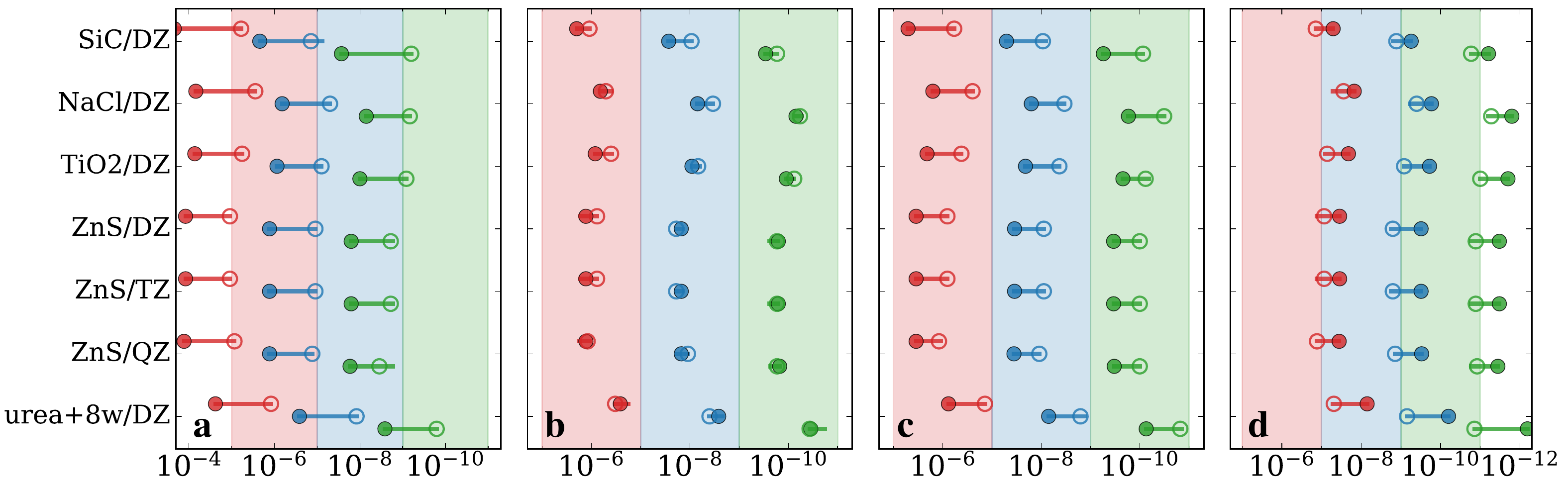}
        \caption{Same plot as Fig.\ M3 for the $O_l v_l$ estimator with different choices of the $f$-prefactor to solve for the cutoffs Eq.\ M51: $f = 1$ (a), $f = \eta_{ab\omega}^{-1}$ (b), $f = R_{\tr{cut}}$ (c), $f = R_{\tr{cut}} \eta_{ab\omega}^{-1}$ (d).
        Note that panel (b) is the same as Fig.\ M2d.}
    \end{figure}

    \begin{figure}[!h]
        \centering
        \includegraphics[width=0.84\linewidth]{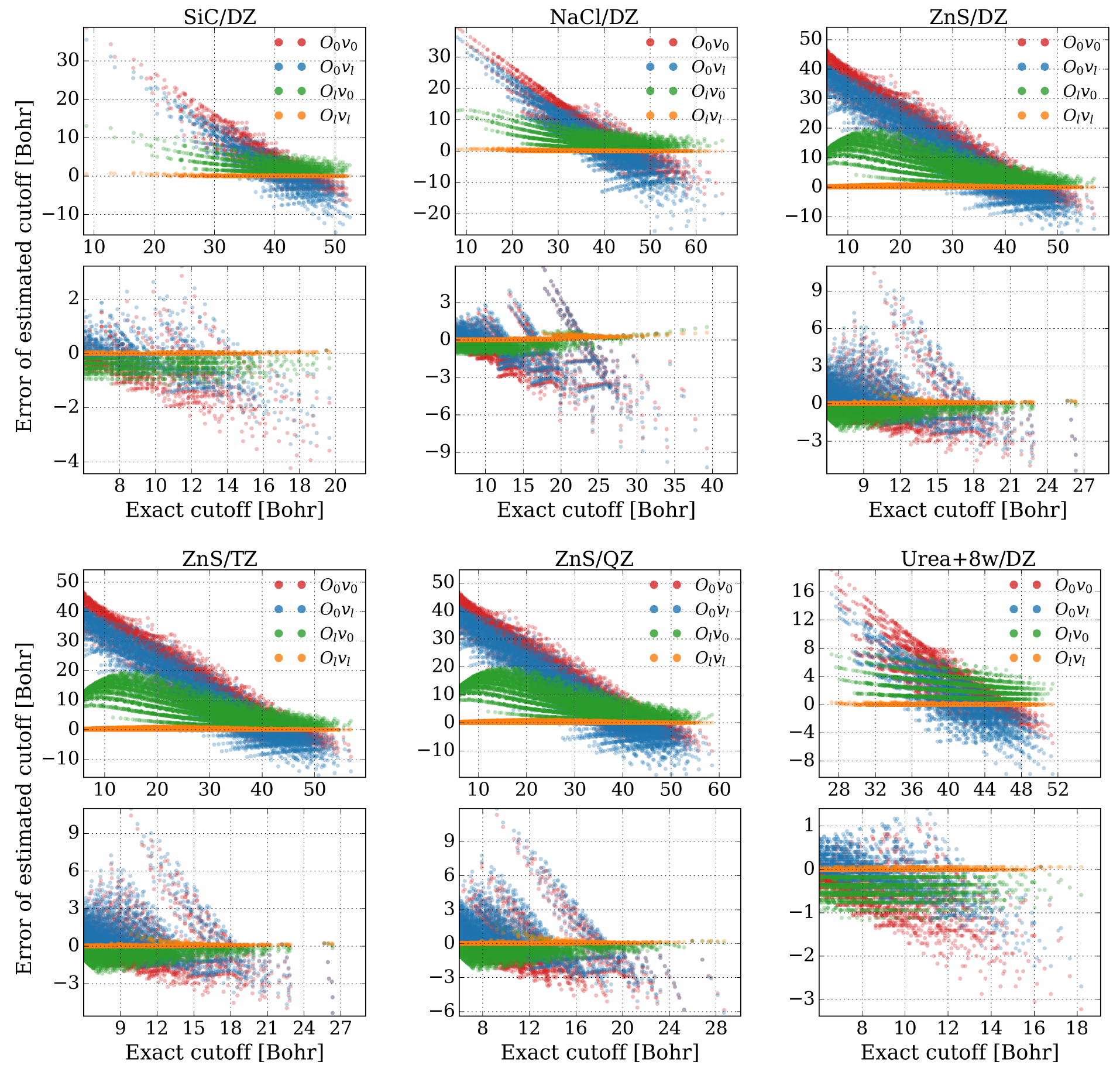}
        \caption{Same plot as Fig.\ M2 for all other systems studied in this work.}
    \end{figure}

    \begin{figure}[!h]
        \centering
        \includegraphics[width=\linewidth]{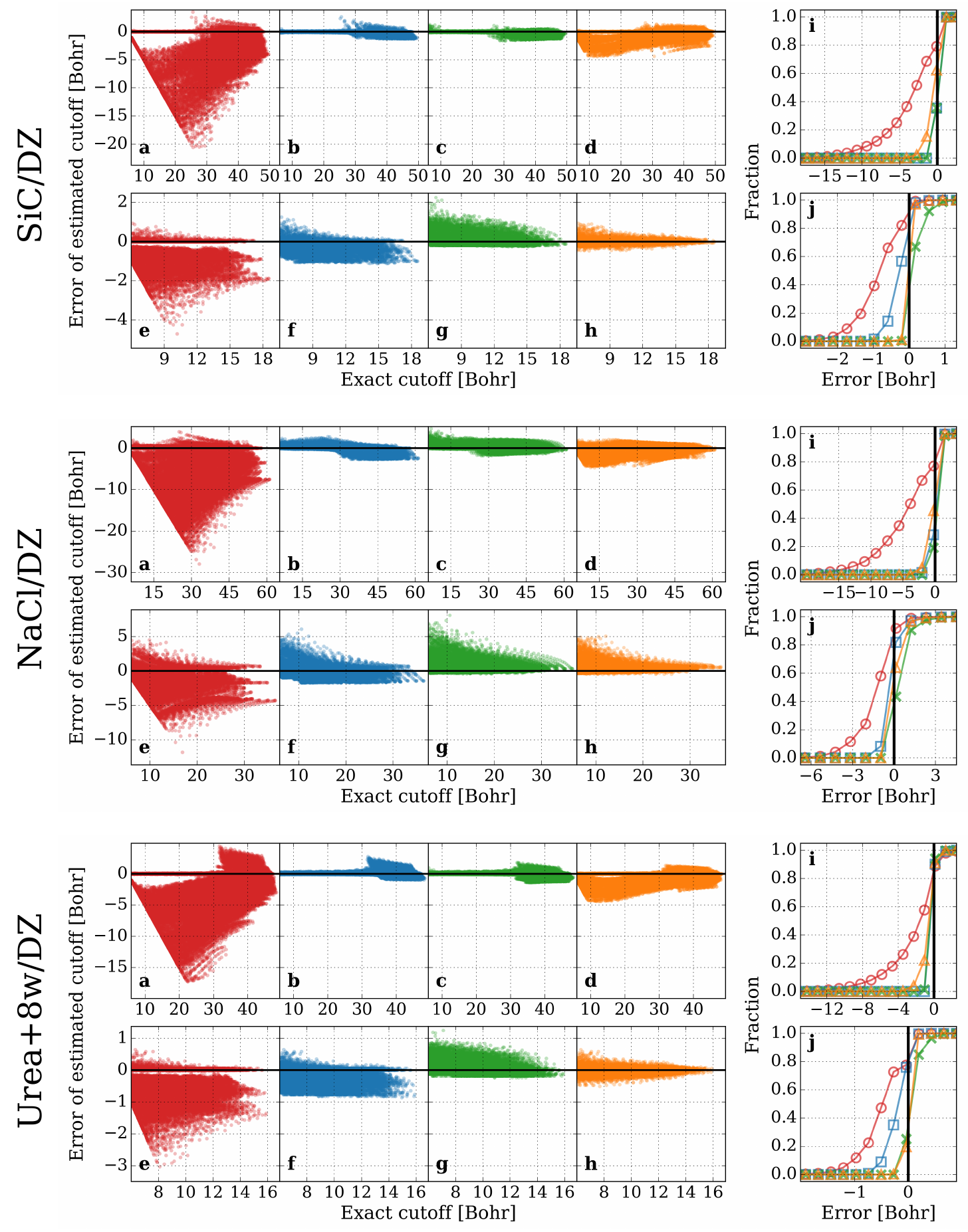}
        \caption{Same plot as Fig.\ M4 for all SiC/DZ, NaCl/DZ, and Ureal+\ce{8H2O}/DZ.}
    \end{figure}

    \begin{figure}[!h]
        \centering
        \includegraphics[width=\linewidth]{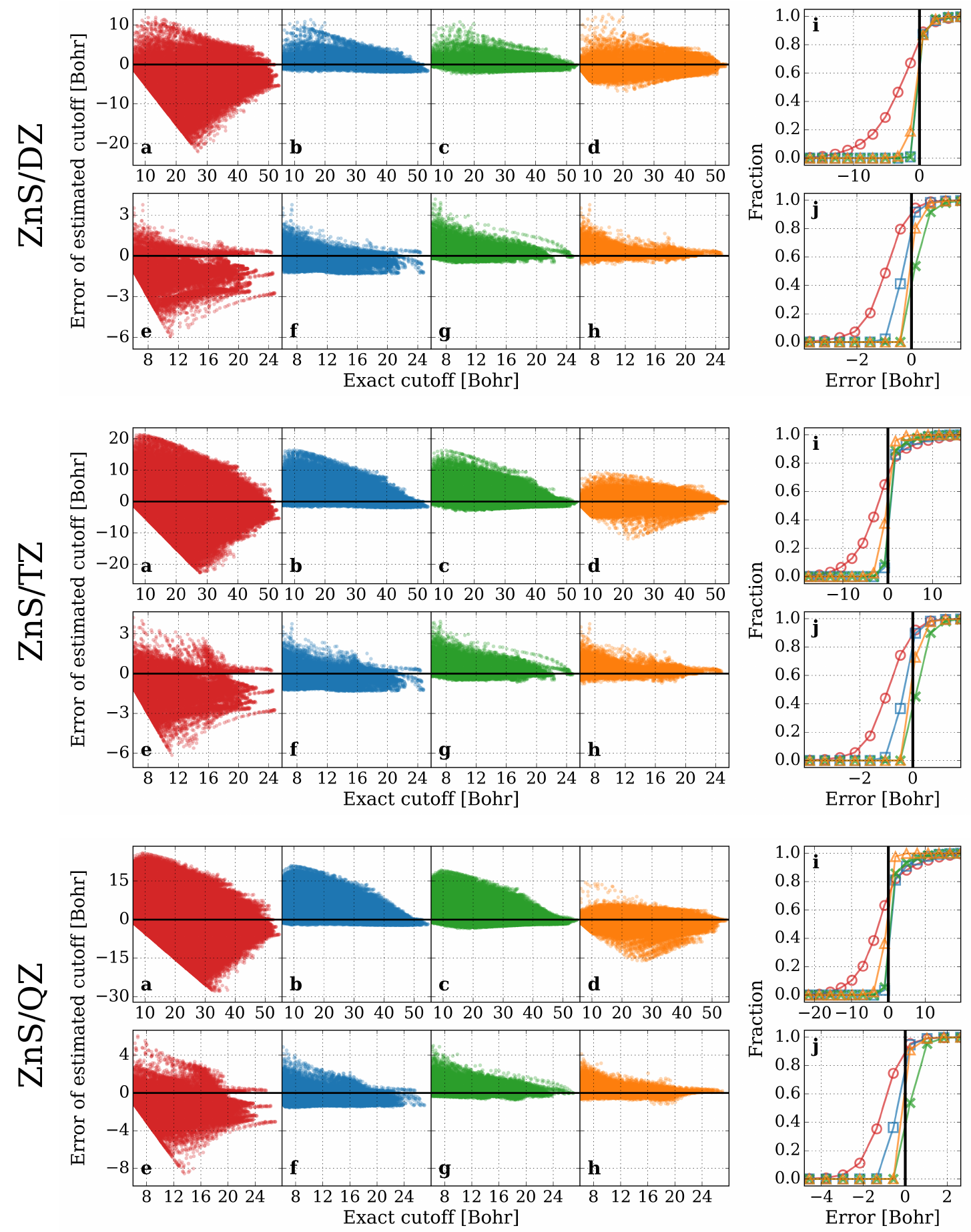}
        \caption{Same plot as Fig.\ M4 for all ZnS/DZ, ZnS/TZ, and ZnS/QZ.}
    \end{figure}

    \begin{figure}[!h]
        \centering
        \includegraphics[width=\linewidth]{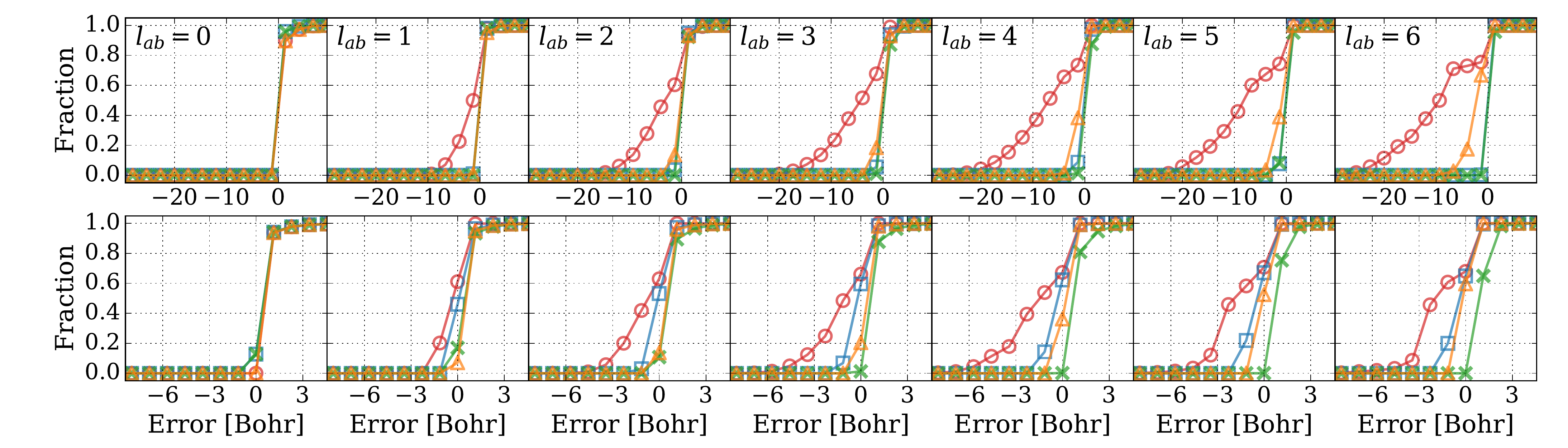}
        \caption{Same plot as Fig.\ M4(i) and (j) for \ce{TiO2}/DZ (upper panels: $\omega = 0.1$; lower panels: $\omega = 1$) except that the data are grouped by the summed orbital angular momentum of the bra AO pairs, $l_{ab}$.}
    \end{figure}

    \begin{figure}[!h]
        \centering
        \includegraphics[width=\linewidth]{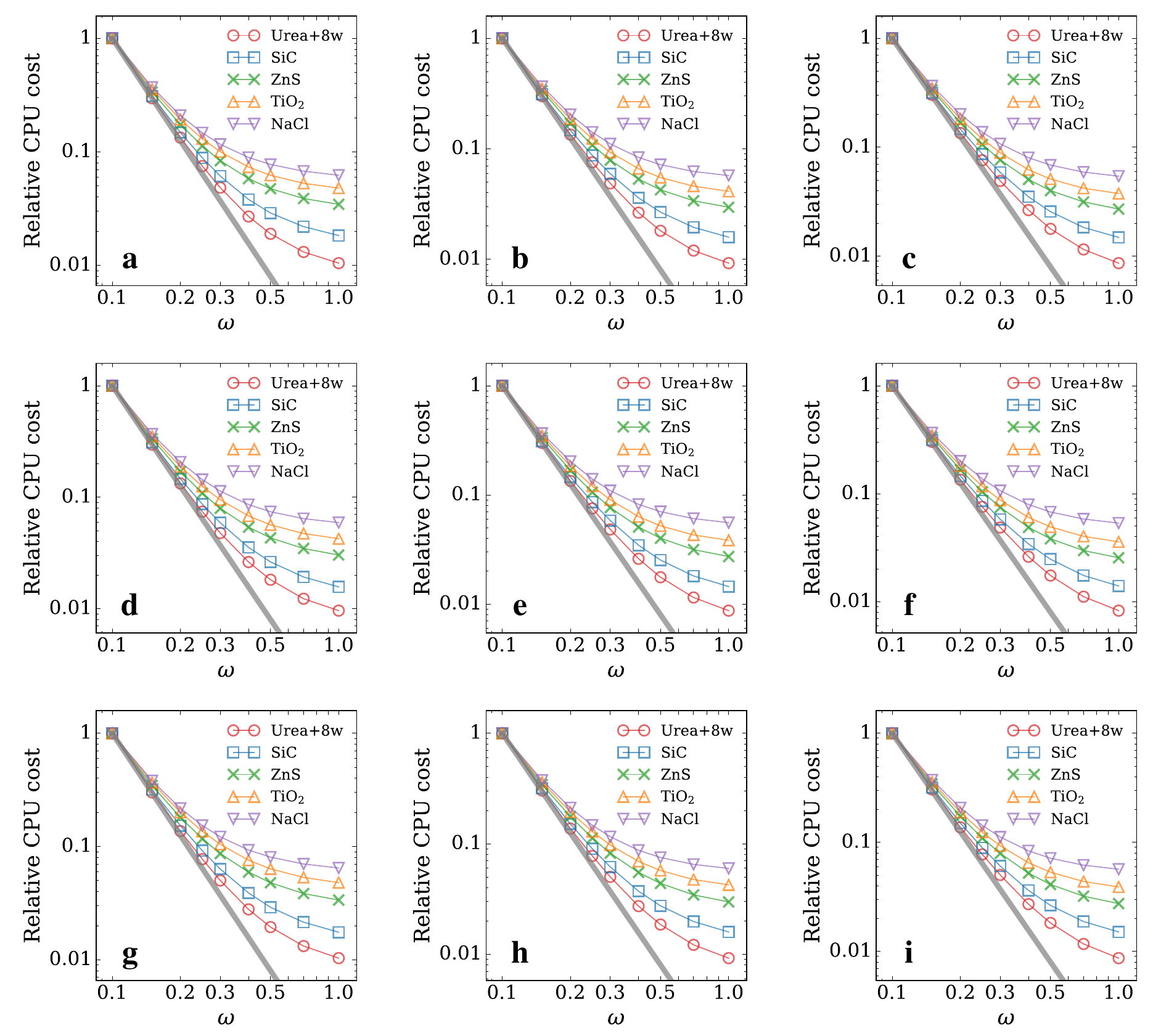}
        \caption{Same plot as Fig.\ M5 but for different choices of estimators and $\epsilon$.
        (a) -- (c) ISF, (d) -- (f) ISF$Q_0$, (g) -- (i) ME.
        For each row, the three panels are generated with $\epsilon = 10^{-6}$, $10^{-8}$, and $10^{-10}$, respectively.
        Note that panel (f) is the same as Fig.\ M5.}
    \end{figure}

    \begin{figure}[!h]
        \centering
        \includegraphics[width=0.4\linewidth]{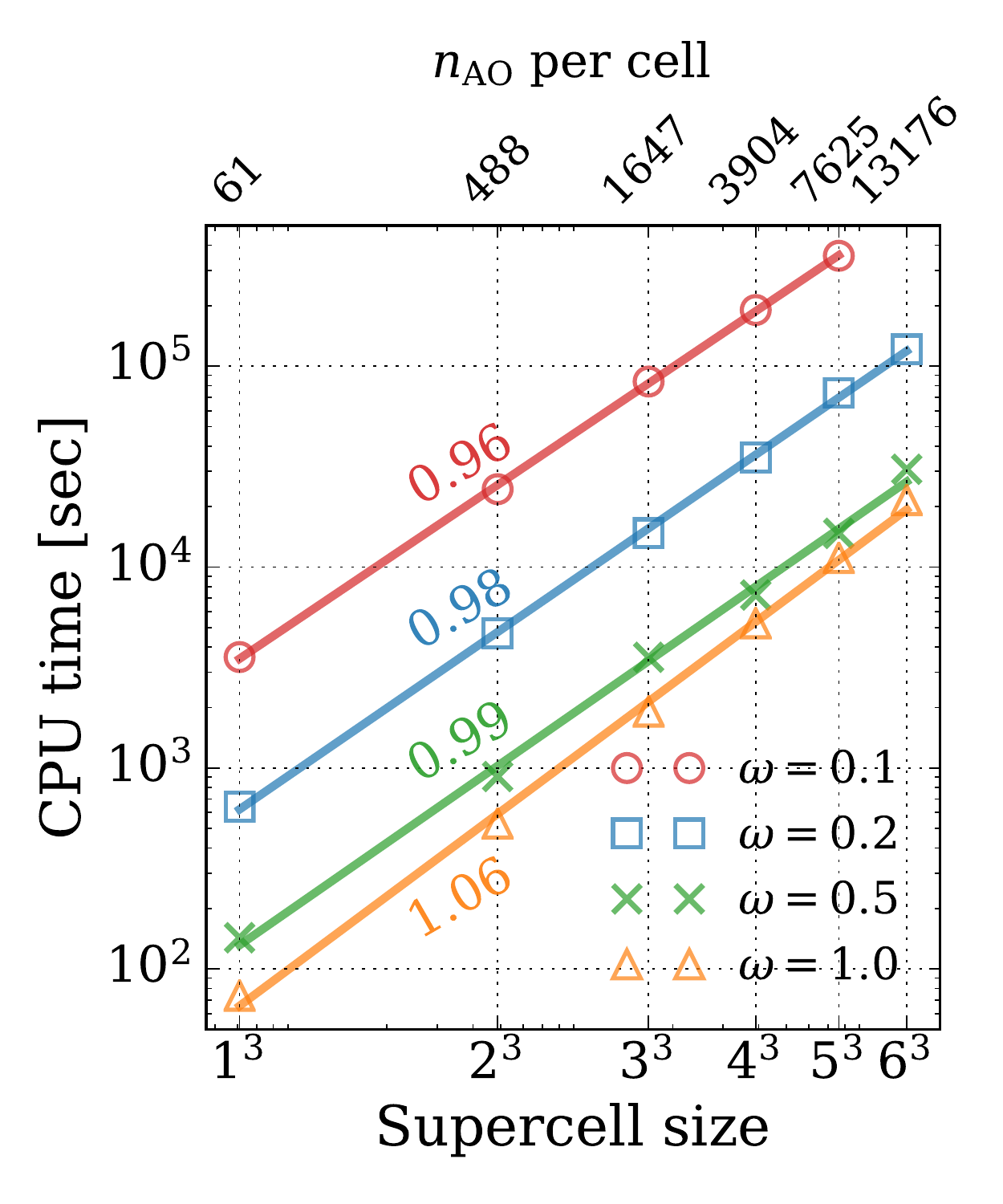}
        \hspace{2em}
        \includegraphics[width=0.4\linewidth]{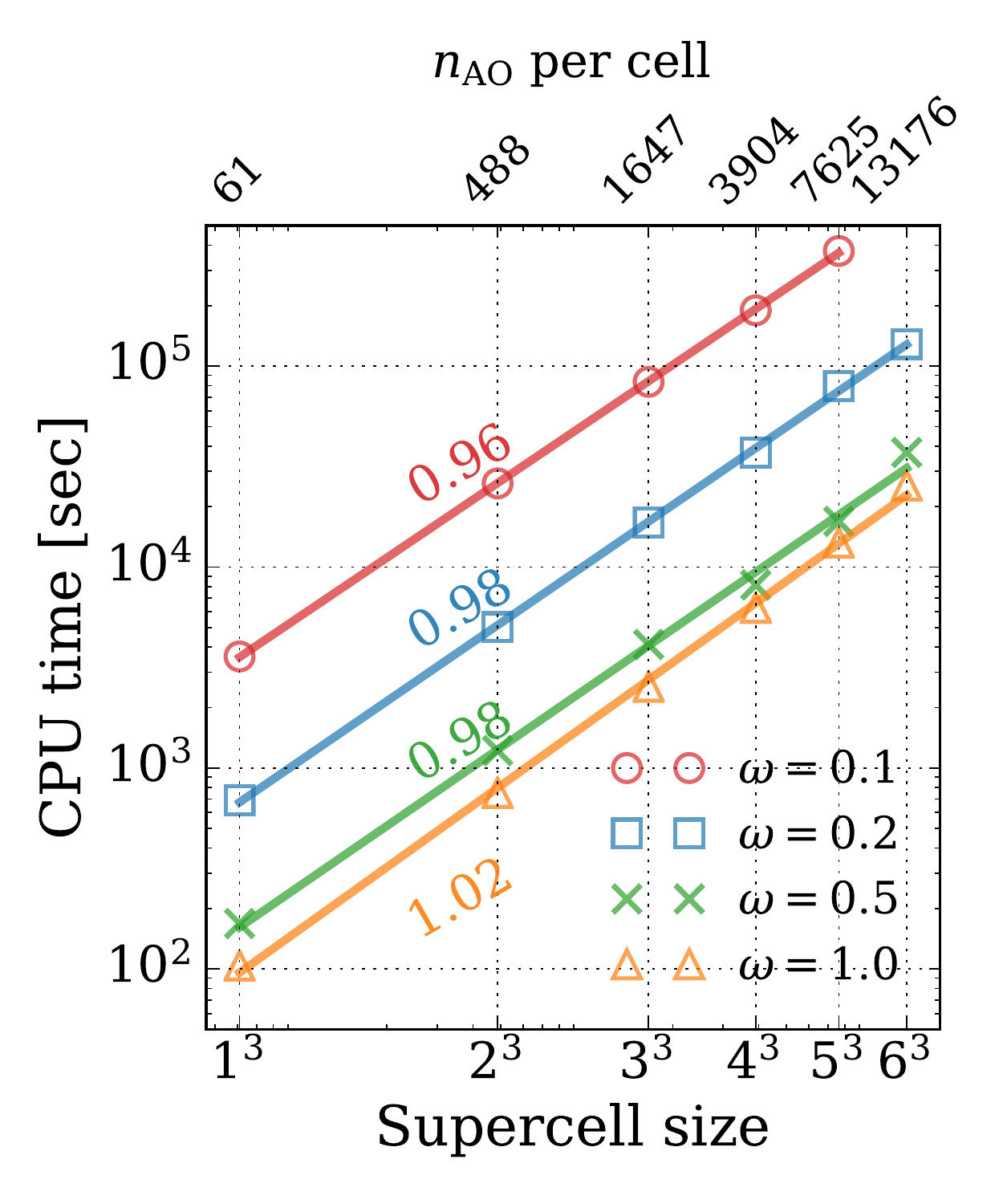}
        \caption{Same plot as Fig.\ M6 except for the ISF$Q_0$ (left) and ISF$Q_l$ (right) estimators.}
    \end{figure}

    \clearpage

    \section{Cartesian coordinates of the test systems in Fig.\ M1}

    The water-solvated urea is optimized at PBE level using Quantum Espresso [1], while others all taken from the Materials Project website.
    The structures are listed below in the format of the VASP POSCAR file.

    \begin{itemize}
        \item SiC:

\begin{Verbatim}
si1 c1
1.0
        3.0968165398         0.0000000000         0.0000000000
        1.5484082699         2.6819217943         0.0000000000
        1.5484082699         0.8939739314         2.5285401165
   Si    C
    1    1
Cartesian
     4.645224571         2.681921721         1.896405101
     0.000000000         0.000000000         0.000000000
\end{Verbatim}

        \item ZnS:

\begin{Verbatim}
zn1 s1
1.0
        3.8539228439         0.0000000000         0.0000000000
        1.9269614220         3.3375950871         0.0000000000
        1.9269614220         1.1125316957         3.1467148252
   Zn    S
    1    1
Cartesian
     0.000000000         0.000000000         0.000000000
     5.780884266         3.337594986         2.360036135
\end{Verbatim}

        \item \ce{TiO2}:

\begin{Verbatim}
ti2 o4
1.0
        4.6532721519         0.0000000000         0.0000000000
        0.0000000000         4.6532721519         0.0000000000
        0.0000000000         0.0000000000         2.9692029953
   Ti    O
    2    4
Cartesian
     2.326636076         2.326636076         1.484601498
     0.000000000         0.000000000         0.000000000
     0.909342408         3.743929625         1.484601498
     3.743929625         0.909342408         1.484601498
     1.417293668         1.417293668         0.000000000
     3.235978603         3.235978603         0.000000000
\end{Verbatim}

        \item \ce{NaCl}:

\begin{Verbatim}
na4 cl4
1.0
        5.6916937828         0.0000000000         0.0000000000
        0.0000000000         5.6916937828         0.0000000000
        0.0000000000         0.0000000000         5.6916937828
   Na   Cl
    4    4
Cartesian
     0.000000000         0.000000000         0.000000000
     0.000000000         2.845846891         2.845846891
     2.845846891         0.000000000         2.845846891
     2.845846891         2.845846891         0.000000000
     2.845846891         0.000000000         0.000000000
     2.845846891         2.845846891         2.845846891
     0.000000000         0.000000000         2.845846891
     0.000000000         2.845846891         0.000000000
\end{Verbatim}

        \item \ce{CO(NH2)2 + 8 H2O}:

\begin{Verbatim}
opt/w8_u1.xyz
1.0
        7.0000100136         0.0000000000         0.0000000000
        0.0000000000         7.0000100136         0.0000000000
        0.0000000000         0.0000000000         7.0000100136
    C    H    N    O
    1   20    2    9
Cartesian
     2.114744902         1.112133145         4.689898968
     5.858695507         5.137805939         1.606225967
     5.772367954         6.110748291         0.300243974
     3.475453854         4.738635063         5.818874359
     4.900005341         4.512650490         6.412111282
     3.027989626         1.066418648         0.408960432
     1.880431294         1.816695333         1.115824580
     4.569515228         0.365846038         5.707927227
     6.117840290         0.606375813         5.531702995
     6.958346844         3.276045322         0.525006950
     6.410978794         1.831803322         0.435900956
     1.661094666         5.845547676         5.571903229
     0.666638434         5.158105373         6.597812653
     4.296371460         1.647681236         2.256358862
     4.984510899         0.560827553         3.078486919
     5.644823551         3.450089455         3.122845411
     5.253567219         4.553706169         4.100537777
     1.679444432         3.125636101         4.936808109
     3.285182714         2.652383327         5.369400978
     0.912300110         6.955224037         3.566261530
     0.455770701         1.570324898         3.542329550
     2.378637075         2.385869026         4.974523067
     0.898828924         0.830556989         4.083635330
     6.066536427         5.223955631         0.640273750
     4.313398838         4.261108875         5.660450935
     2.760848999         1.393758059         1.285377860
     5.426894188         0.447687060         6.206815720
     0.239478484         2.344601870         0.525727928
     1.394407511         4.987095833         5.961184502
     5.110028267         1.480589747         2.791260481
     5.828163147         4.399785519         3.308534861
     2.892082214         0.158870816         4.997205257
\end{Verbatim}
    \end{itemize}

    \clearpage

    \section{Even tempered basis (ETB) used in this work}

    Generated by PySCF with the following code (Zn/cc-pVDZ as an example):
    \begin{verbatim}
        from pyscf import gto, df
        mol = gto.M(atom="Zn", basis="cc-pVDZ", spin=None)
        auxmol = df.make_auxmol(mol)
        print(auxmol._basis["Zn"])  # ETB with progression beta = 2.0
    \end{verbatim}

\begin{Verbatim}
# Na: cc-pVDZ/ETB
Na S
6047.6620800000 1.0000000000 
Na S
3023.8310400000 1.0000000000 
Na S
1511.9155200000 1.0000000000 
Na S
755.9577600000 1.0000000000 
Na S
377.9788800000 1.0000000000 
Na S
188.9894400000 1.0000000000 
Na S
94.4947200000 1.0000000000 
Na S
47.2473600000 1.0000000000 
Na S
23.6236800000 1.0000000000 
Na S
11.8118400000 1.0000000000 
Na S
5.9059200000 1.0000000000 
Na S
2.9529600000 1.0000000000 
Na S
1.4764800000 1.0000000000 
Na S
0.7382400000 1.0000000000 
Na S
0.3691200000 1.0000000000 
Na S
0.1845600000 1.0000000000 
Na S
0.0922800000 1.0000000000 
Na S
0.0461400000 1.0000000000 
Na P
1426.2583584328 1.0000000000 
Na P
713.1291792164 1.0000000000 
Na P
356.5645896082 1.0000000000 
Na P
178.2822948041 1.0000000000 
Na P
89.1411474020 1.0000000000 
Na P
44.5705737010 1.0000000000 
Na P
22.2852868505 1.0000000000 
Na P
11.1426434253 1.0000000000 
Na P
5.5713217126 1.0000000000 
Na P
2.7856608563 1.0000000000 
Na P
1.3928304282 1.0000000000 
Na P
0.6964152141 1.0000000000 
Na P
0.3482076070 1.0000000000 
Na P
0.1741038035 1.0000000000 
Na P
0.0870519018 1.0000000000 
Na P
0.0435259509 1.0000000000 
Na D
168.1817600000 1.0000000000 
Na D
84.0908800000 1.0000000000 
Na D
42.0454400000 1.0000000000 
Na D
21.0227200000 1.0000000000 
Na D
10.5113600000 1.0000000000 
Na D
5.2556800000 1.0000000000 
Na D
2.6278400000 1.0000000000 
Na D
1.3139200000 1.0000000000 
Na D
0.6569600000 1.0000000000 
Na D
0.3284800000 1.0000000000 
Na D
0.1642400000 1.0000000000 
Na D
0.0821200000 1.0000000000 
Na D
0.0410600000 1.0000000000 
Na F
5.7208540006 1.0000000000 
Na F
2.8604270003 1.0000000000 
Na F
1.4302135001 1.0000000000 
Na F
0.7151067501 1.0000000000 
Na F
0.3575533750 1.0000000000 
Na F
0.1787766875 1.0000000000 
Na F
0.0893883438 1.0000000000 
Na G
0.1946000000 1.0000000000 
\end{Verbatim}
\begin{Verbatim}
# Ti: cc-pVDZ/ETB
Ti S
53403.9756800000 1.0000000000 
Ti S
26701.9878400000 1.0000000000 
Ti S
13350.9939200000 1.0000000000 
Ti S
6675.4969600000 1.0000000000 
Ti S
3337.7484800000 1.0000000000 
Ti S
1668.8742400000 1.0000000000 
Ti S
834.4371200000 1.0000000000 
Ti S
417.2185600000 1.0000000000 
Ti S
208.6092800000 1.0000000000 
Ti S
104.3046400000 1.0000000000 
Ti S
52.1523200000 1.0000000000 
Ti S
26.0761600000 1.0000000000 
Ti S
13.0380800000 1.0000000000 
Ti S
6.5190400000 1.0000000000 
Ti S
3.2595200000 1.0000000000 
Ti S
1.6297600000 1.0000000000 
Ti S
0.8148800000 1.0000000000 
Ti S
0.4074400000 1.0000000000 
Ti S
0.2037200000 1.0000000000 
Ti S
0.1018600000 1.0000000000 
Ti S
0.0509300000 1.0000000000 
Ti P
7172.8675792108 1.0000000000 
Ti P
3586.4337896054 1.0000000000 
Ti P
1793.2168948027 1.0000000000 
Ti P
896.6084474013 1.0000000000 
Ti P
448.3042237007 1.0000000000 
Ti P
224.1521118503 1.0000000000 
Ti P
112.0760559252 1.0000000000 
Ti P
56.0380279626 1.0000000000 
Ti P
28.0190139813 1.0000000000 
Ti P
14.0095069906 1.0000000000 
Ti P
7.0047534953 1.0000000000 
Ti P
3.5023767477 1.0000000000 
Ti P
1.7511883738 1.0000000000 
Ti P
0.8755941869 1.0000000000 
Ti P
0.4377970935 1.0000000000 
Ti P
0.2188985467 1.0000000000 
Ti P
0.1094492734 1.0000000000 
Ti P
0.0547246367 1.0000000000 
Ti D
1926.8239360000 1.0000000000 
Ti D
963.4119680000 1.0000000000 
Ti D
481.7059840000 1.0000000000 
Ti D
240.8529920000 1.0000000000 
Ti D
120.4264960000 1.0000000000 
Ti D
60.2132480000 1.0000000000 
Ti D
30.1066240000 1.0000000000 
Ti D
15.0533120000 1.0000000000 
Ti D
7.5266560000 1.0000000000 
Ti D
3.7633280000 1.0000000000 
Ti D
1.8816640000 1.0000000000 
Ti D
0.9408320000 1.0000000000 
Ti D
0.4704160000 1.0000000000 
Ti D
0.2352080000 1.0000000000 
Ti D
0.1176040000 1.0000000000 
Ti D
0.0588020000 1.0000000000 
Ti F
319.9541296650 1.0000000000 
Ti F
159.9770648325 1.0000000000 
Ti F
79.9885324163 1.0000000000 
Ti F
39.9942662081 1.0000000000 
Ti F
19.9971331041 1.0000000000 
Ti F
9.9985665520 1.0000000000 
Ti F
4.9992832760 1.0000000000 
Ti F
2.4996416380 1.0000000000 
Ti F
1.2498208190 1.0000000000 
Ti F
0.6249104095 1.0000000000 
Ti F
0.3124552048 1.0000000000 
Ti F
0.1562276024 1.0000000000 
Ti F
0.0781138012 1.0000000000 
Ti G
106.2584320000 1.0000000000 
Ti G
53.1292160000 1.0000000000 
Ti G
26.5646080000 1.0000000000 
Ti G
13.2823040000 1.0000000000 
Ti G
6.6411520000 1.0000000000 
Ti G
3.3205760000 1.0000000000 
Ti G
1.6602880000 1.0000000000 
Ti G
0.8301440000 1.0000000000 
Ti G
0.4150720000 1.0000000000 
Ti G
0.2075360000 1.0000000000 
Ti G
0.1037680000 1.0000000000 
Ti H
15.3947655628 1.0000000000 
Ti H
7.6973827814 1.0000000000 
Ti H
3.8486913907 1.0000000000 
Ti H
1.9243456953 1.0000000000 
Ti H
0.9621728477 1.0000000000 
Ti H
0.4810864238 1.0000000000 
Ti H
0.2405432119 1.0000000000 
Ti I
2.2304000000 1.0000000000 
Ti I
1.1152000000 1.0000000000 
Ti I
0.5576000000 1.0000000000 
\end{Verbatim}
\begin{Verbatim}
# Zn: cc-pVDZ/ETB
Zn S
79066.8247040000 1.0000000000 
Zn S
39533.4123520000 1.0000000000 
Zn S
19766.7061760000 1.0000000000 
Zn S
9883.3530880000 1.0000000000 
Zn S
4941.6765440000 1.0000000000 
Zn S
2470.8382720000 1.0000000000 
Zn S
1235.4191360000 1.0000000000 
Zn S
617.7095680000 1.0000000000 
Zn S
308.8547840000 1.0000000000 
Zn S
154.4273920000 1.0000000000 
Zn S
77.2136960000 1.0000000000 
Zn S
38.6068480000 1.0000000000 
Zn S
19.3034240000 1.0000000000 
Zn S
9.6517120000 1.0000000000 
Zn S
4.8258560000 1.0000000000 
Zn S
2.4129280000 1.0000000000 
Zn S
1.2064640000 1.0000000000 
Zn S
0.6032320000 1.0000000000 
Zn S
0.3016160000 1.0000000000 
Zn S
0.1508080000 1.0000000000 
Zn S
0.0754040000 1.0000000000 
Zn P
10950.3544120812 1.0000000000 
Zn P
5475.1772060406 1.0000000000 
Zn P
2737.5886030203 1.0000000000 
Zn P
1368.7943015102 1.0000000000 
Zn P
684.3971507551 1.0000000000 
Zn P
342.1985753775 1.0000000000 
Zn P
171.0992876888 1.0000000000 
Zn P
85.5496438444 1.0000000000 
Zn P
42.7748219222 1.0000000000 
Zn P
21.3874109611 1.0000000000 
Zn P
10.6937054805 1.0000000000 
Zn P
5.3468527403 1.0000000000 
Zn P
2.6734263701 1.0000000000 
Zn P
1.3367131851 1.0000000000 
Zn P
0.6683565925 1.0000000000 
Zn P
0.3341782963 1.0000000000 
Zn P
0.1670891481 1.0000000000 
Zn P
0.0835445741 1.0000000000 
Zn D
6066.2743040000 1.0000000000 
Zn D
3033.1371520000 1.0000000000 
Zn D
1516.5685760000 1.0000000000 
Zn D
758.2842880000 1.0000000000 
Zn D
379.1421440000 1.0000000000 
Zn D
189.5710720000 1.0000000000 
Zn D
94.7855360000 1.0000000000 
Zn D
47.3927680000 1.0000000000 
Zn D
23.6963840000 1.0000000000 
Zn D
11.8481920000 1.0000000000 
Zn D
5.9240960000 1.0000000000 
Zn D
2.9620480000 1.0000000000 
Zn D
1.4810240000 1.0000000000 
Zn D
0.7405120000 1.0000000000 
Zn D
0.3702560000 1.0000000000 
Zn D
0.1851280000 1.0000000000 
Zn D
0.0925640000 1.0000000000 
Zn F
884.0121324288 1.0000000000 
Zn F
442.0060662144 1.0000000000 
Zn F
221.0030331072 1.0000000000 
Zn F
110.5015165536 1.0000000000 
Zn F
55.2507582768 1.0000000000 
Zn F
27.6253791384 1.0000000000 
Zn F
13.8126895692 1.0000000000 
Zn F
6.9063447846 1.0000000000 
Zn F
3.4531723923 1.0000000000 
Zn F
1.7265861962 1.0000000000 
Zn F
0.8632930981 1.0000000000 
Zn F
0.4316465490 1.0000000000 
Zn F
0.2158232745 1.0000000000 
Zn G
257.6465920000 1.0000000000 
Zn G
128.8232960000 1.0000000000 
Zn G
64.4116480000 1.0000000000 
Zn G
32.2058240000 1.0000000000 
Zn G
16.1029120000 1.0000000000 
Zn G
8.0514560000 1.0000000000 
Zn G
4.0257280000 1.0000000000 
Zn G
2.0128640000 1.0000000000 
Zn G
1.0064320000 1.0000000000 
Zn G
0.5032160000 1.0000000000 
Zn H
38.8098180469 1.0000000000 
Zn H
19.4049090234 1.0000000000 
Zn H
9.7024545117 1.0000000000 
Zn H
4.8512272559 1.0000000000 
Zn H
2.4256136279 1.0000000000 
Zn H
1.2128068140 1.0000000000 
Zn I
11.6920000000 1.0000000000 
Zn I
5.8460000000 1.0000000000 
Zn I
2.9230000000 1.0000000000 
\end{Verbatim}
\begin{Verbatim}
# Zn: cc-pVTZ/ETB
Zn S
79066.8247040000 1.0000000000 
Zn S
39533.4123520000 1.0000000000 
Zn S
19766.7061760000 1.0000000000 
Zn S
9883.3530880000 1.0000000000 
Zn S
4941.6765440000 1.0000000000 
Zn S
2470.8382720000 1.0000000000 
Zn S
1235.4191360000 1.0000000000 
Zn S
617.7095680000 1.0000000000 
Zn S
308.8547840000 1.0000000000 
Zn S
154.4273920000 1.0000000000 
Zn S
77.2136960000 1.0000000000 
Zn S
38.6068480000 1.0000000000 
Zn S
19.3034240000 1.0000000000 
Zn S
9.6517120000 1.0000000000 
Zn S
4.8258560000 1.0000000000 
Zn S
2.4129280000 1.0000000000 
Zn S
1.2064640000 1.0000000000 
Zn S
0.6032320000 1.0000000000 
Zn S
0.3016160000 1.0000000000 
Zn S
0.1508080000 1.0000000000 
Zn S
0.0754040000 1.0000000000 
Zn P
10950.3544120812 1.0000000000 
Zn P
5475.1772060406 1.0000000000 
Zn P
2737.5886030203 1.0000000000 
Zn P
1368.7943015102 1.0000000000 
Zn P
684.3971507551 1.0000000000 
Zn P
342.1985753775 1.0000000000 
Zn P
171.0992876888 1.0000000000 
Zn P
85.5496438444 1.0000000000 
Zn P
42.7748219222 1.0000000000 
Zn P
21.3874109611 1.0000000000 
Zn P
10.6937054805 1.0000000000 
Zn P
5.3468527403 1.0000000000 
Zn P
2.6734263701 1.0000000000 
Zn P
1.3367131851 1.0000000000 
Zn P
0.6683565925 1.0000000000 
Zn P
0.3341782963 1.0000000000 
Zn P
0.1670891481 1.0000000000 
Zn P
0.0835445741 1.0000000000 
Zn D
6066.2743040000 1.0000000000 
Zn D
3033.1371520000 1.0000000000 
Zn D
1516.5685760000 1.0000000000 
Zn D
758.2842880000 1.0000000000 
Zn D
379.1421440000 1.0000000000 
Zn D
189.5710720000 1.0000000000 
Zn D
94.7855360000 1.0000000000 
Zn D
47.3927680000 1.0000000000 
Zn D
23.6963840000 1.0000000000 
Zn D
11.8481920000 1.0000000000 
Zn D
5.9240960000 1.0000000000 
Zn D
2.9620480000 1.0000000000 
Zn D
1.4810240000 1.0000000000 
Zn D
0.7405120000 1.0000000000 
Zn D
0.3702560000 1.0000000000 
Zn D
0.1851280000 1.0000000000 
Zn D
0.0925640000 1.0000000000 
Zn F
884.0121324288 1.0000000000 
Zn F
442.0060662144 1.0000000000 
Zn F
221.0030331072 1.0000000000 
Zn F
110.5015165536 1.0000000000 
Zn F
55.2507582768 1.0000000000 
Zn F
27.6253791384 1.0000000000 
Zn F
13.8126895692 1.0000000000 
Zn F
6.9063447846 1.0000000000 
Zn F
3.4531723923 1.0000000000 
Zn F
1.7265861962 1.0000000000 
Zn F
0.8632930981 1.0000000000 
Zn F
0.4316465490 1.0000000000 
Zn F
0.2158232745 1.0000000000 
Zn G
257.6465920000 1.0000000000 
Zn G
128.8232960000 1.0000000000 
Zn G
64.4116480000 1.0000000000 
Zn G
32.2058240000 1.0000000000 
Zn G
16.1029120000 1.0000000000 
Zn G
8.0514560000 1.0000000000 
Zn G
4.0257280000 1.0000000000 
Zn G
2.0128640000 1.0000000000 
Zn G
1.0064320000 1.0000000000 
Zn G
0.5032160000 1.0000000000 
Zn H
39.1219096558 1.0000000000 
Zn H
19.5609548279 1.0000000000 
Zn H
9.7804774140 1.0000000000 
Zn H
4.8902387070 1.0000000000 
Zn H
2.4451193535 1.0000000000 
Zn H
1.2225596767 1.0000000000 
Zn I
11.8808000000 1.0000000000 
Zn I
5.9404000000 1.0000000000 
Zn I
2.9702000000 1.0000000000 
\end{Verbatim}
\begin{Verbatim}
# Zn: cc-pVQZ/ETB
Zn S
77135.3477120000 1.0000000000 
Zn S
38567.6738560000 1.0000000000 
Zn S
19283.8369280000 1.0000000000 
Zn S
9641.9184640000 1.0000000000 
Zn S
4820.9592320000 1.0000000000 
Zn S
2410.4796160000 1.0000000000 
Zn S
1205.2398080000 1.0000000000 
Zn S
602.6199040000 1.0000000000 
Zn S
301.3099520000 1.0000000000 
Zn S
150.6549760000 1.0000000000 
Zn S
75.3274880000 1.0000000000 
Zn S
37.6637440000 1.0000000000 
Zn S
18.8318720000 1.0000000000 
Zn S
9.4159360000 1.0000000000 
Zn S
4.7079680000 1.0000000000 
Zn S
2.3539840000 1.0000000000 
Zn S
1.1769920000 1.0000000000 
Zn S
0.5884960000 1.0000000000 
Zn S
0.2942480000 1.0000000000 
Zn S
0.1471240000 1.0000000000 
Zn S
0.0735620000 1.0000000000 
Zn P
9897.2732086246 1.0000000000 
Zn P
4948.6366043123 1.0000000000 
Zn P
2474.3183021561 1.0000000000 
Zn P
1237.1591510781 1.0000000000 
Zn P
618.5795755390 1.0000000000 
Zn P
309.2897877695 1.0000000000 
Zn P
154.6448938848 1.0000000000 
Zn P
77.3224469424 1.0000000000 
Zn P
38.6612234712 1.0000000000 
Zn P
19.3306117356 1.0000000000 
Zn P
9.6653058678 1.0000000000 
Zn P
4.8326529339 1.0000000000 
Zn P
2.4163264669 1.0000000000 
Zn P
1.2081632335 1.0000000000 
Zn P
0.6040816167 1.0000000000 
Zn P
0.3020408084 1.0000000000 
Zn P
0.1510204042 1.0000000000 
Zn P
0.0755102021 1.0000000000 
Zn D
5079.6953600000 1.0000000000 
Zn D
2539.8476800000 1.0000000000 
Zn D
1269.9238400000 1.0000000000 
Zn D
634.9619200000 1.0000000000 
Zn D
317.4809600000 1.0000000000 
Zn D
158.7404800000 1.0000000000 
Zn D
79.3702400000 1.0000000000 
Zn D
39.6851200000 1.0000000000 
Zn D
19.8425600000 1.0000000000 
Zn D
9.9212800000 1.0000000000 
Zn D
4.9606400000 1.0000000000 
Zn D
2.4803200000 1.0000000000 
Zn D
1.2401600000 1.0000000000 
Zn D
0.6200800000 1.0000000000 
Zn D
0.3100400000 1.0000000000 
Zn D
0.1550200000 1.0000000000 
Zn D
0.0775100000 1.0000000000 
Zn F
1287.4848516749 1.0000000000 
Zn F
643.7424258374 1.0000000000 
Zn F
321.8712129187 1.0000000000 
Zn F
160.9356064594 1.0000000000 
Zn F
80.4678032297 1.0000000000 
Zn F
40.2339016148 1.0000000000 
Zn F
20.1169508074 1.0000000000 
Zn F
10.0584754037 1.0000000000 
Zn F
5.0292377019 1.0000000000 
Zn F
2.5146188509 1.0000000000 
Zn F
1.2573094255 1.0000000000 
Zn F
0.6286547127 1.0000000000 
Zn F
0.3143273564 1.0000000000 
Zn F
0.1571636782 1.0000000000 
Zn G
326.3221760000 1.0000000000 
Zn G
163.1610880000 1.0000000000 
Zn G
81.5805440000 1.0000000000 
Zn G
40.7902720000 1.0000000000 
Zn G
20.3951360000 1.0000000000 
Zn G
10.1975680000 1.0000000000 
Zn G
5.0987840000 1.0000000000 
Zn G
2.5493920000 1.0000000000 
Zn G
1.2746960000 1.0000000000 
Zn G
0.6373480000 1.0000000000 
Zn G
0.3186740000 1.0000000000 
Zn H
47.1892204748 1.0000000000 
Zn H
23.5946102374 1.0000000000 
Zn H
11.7973051187 1.0000000000 
Zn H
5.8986525594 1.0000000000 
Zn H
2.9493262797 1.0000000000 
Zn H
1.4746631398 1.0000000000 
Zn H
0.7373315699 1.0000000000 
Zn I
13.6480000000 1.0000000000 
Zn I
6.8240000000 1.0000000000 
Zn I
3.4120000000 1.0000000000 
Zn I
1.7060000000 1.0000000000 
\end{Verbatim}


    \clearpage

    \section*{References}

    \begin{enumerate}
        \item [(1)] Paolo Giannozzi, Oscar Baseggio, Pietro Bonf\`{a}, Davide Brunato, Roberto Car, Ivan Carnimeo, Carlo Cavazzoni, Stefano de Gironcoli, Pietro Delugas, Fabrizio Ferrari Ruffino, Andrea Ferretti, Nicola Marzari, Iurii Timrov, Andrea Urru, and Stefano Baroni , "Quantum ESPRESSO toward the exascale", \textit{The Journal of Chemical Physics} \textbf{152}, 154105 (2020) https://doi.org/10.1063/5.0005082
    \end{enumerate}